\pdfoutput=1
\documentclass[preprint,12pt]{elsarticle}

\usepackage[top=2.0cm,bottom=2.0cm,left=2.0cm,right=2.0cm]{geometry}
\usepackage{float}
\usepackage{graphicx}
\usepackage{setspace}
\usepackage{amssymb}
\usepackage{subfigure}
\usepackage{caption}
\usepackage{hyperref}
\usepackage{siunitx}
\usepackage{psfrag}
\usepackage{pstool}
\usepackage{changes}
\usepackage{commath}
\usepackage[normalem]{ulem}
\usepackage[numbers]{natbib}
\usepackage{placeins}
\usepackage[symbol]{footmisc}
\usepackage{bm}
\usepackage{amsmath}
\usepackage{chngcntr}
\usepackage{color}
\usepackage{soul}
\usepackage{multirow}
\usepackage[tableposition=top]{caption}
\usepackage{lineno}

\begin{document}

\begin{frontmatter}

\title{Cold dwell behaviour of Ti6Al alloy: Understanding load shedding using digital image correlation and crystal plasticity simulations}

\author[add1]{Yi Xiong\corref{*}}
\cortext[*]{Corresponding author}
\address[add1]{Department of Materials, University of Oxford, Parks Road, Oxford, OX1 3PH, United Kingdom}
\ead{yi.xiong@materials.ox.ac.uk}
\author[add2]{Nicol\`{o} Grilli}
\address[add2]{Department of Engineering Science, University of Oxford, Parks Road, Oxford, OX1 3PJ, United Kingdom}
\author[add1]{Phani S. Karamched}
\author[add1]{Bo-Shiuan Li}
\author[add2,add1]{Edmund Tarleton}
\author[add1]{Angus J. Wilkinson}

\begin{abstract}
Digital image correlation (DIC) and crystal plasticity simulation were utilised to study cold dwell behaviour in a coarse grain Ti-6Al alloy at 3 different temperatures up to 230~$^{\circ}$C. Strains extracted from large volume grains were measured during creep by DIC and were used to calibrate the crystal plasticity model. The values of critical resolved shear stresses (CRSS) of the two main slip systems (basal and prismatic) were determined as a function of temperature. Stress along paths across the boundaries of two grain pairs, (1) a ‘rogue’ grain pair and (2) a ‘non-rogue’ grain pair, were determined at different temperatures. Load shedding was observed in the ‘rogue’ grain pair, where a stress increment during the creep period was found in the ‘hard’ grain. At elevated temperatures, 120~$^{\circ}$C was found to be the worst case scenario as the stress difference at the grain boundaries of these two grain pairs were found to be the largest among the three temperatures. This can be attributed to the fact that the strain rate sensitivity of both prismatic and basal slip systems is at its greatest in this worst case scenario temperature.
\end{abstract}


\end{frontmatter}
\section{Introduction}
\label{Intro}
Titanium alloys with predominantly $\alpha$-phase (HCP crystalline structure) are widely used in aero engines and gas turbines at moderate temperatures, due to their excellent specific strength and corrosion resistance~\cite{ANAHID2011}. In such applications, titanium alloy components often subjected to extremely complicated stress state. For instance, typical flight cycles comprise of takeoff (load up), cruise (stress hold) and landing (load off), combined with a varying temperature from ambient conditions and engine heating. During the cruise phase, titanium fan blades and compressor disks experience long hours of high stress hold (dwell), which often results in a drastic lifetime reduction for these titanium components~\cite{BACHE2003,TYMPEL2016}. This phenomenon that seems to happen at relatively lower temperatures (typically below 200~$^{\circ}$C) has been termed ‘cold dwell fatigue’.

Due to its significance in the aerospace industry, several efforts have been made to study this complicated problem. Cold dwell fatigue mainly occurs in $\alpha$-Ti alloys~\cite{OZTURK2017}. The deformation of HCP crystal structure is highly anisotropic due to its low symmetric elasticity and different slip system strengths~\cite{HASIJA2003,DEKA2006}. During a stress dwell, local time dependent plasticity could occur in grains well orientated for basal and prismatic slip (‘soft’ grains). This leads to load shedding onto grains badly orientated for slip (i.e. ‘hard’ grains with crystallographic $c$-axis of the $\alpha$-Ti grain nearly parallel to the loading direction)~\cite{XU2020}. The ‘hard’ oriented grain, due to higher elastic modulus, encounters a stress increase within the grain. Dunne \emph{et al}~\cite{DUNNE20071061,DUNNE2007Proceedings} used a crystal plasticity finite element (CPFE) model to investigate the worst case scenario, in a combination of a ‘soft’ and ‘hard’ grain, also known as ‘rogue’ grain pair. Upon deformation, large stresses accumulated at the grain boundary and also into the 'hard' grain. Sinha \emph{et al}~\cite{Sinha2004,Sinha2006_1,Sinha2006_2} suggested that dwell fatigue crack nucleation is usually related to \{0002\} facets which lie almost perpendicular to the loading direction. This facet formation within the ‘hard’ grain is initiated by dislocation pile-up in the adjacent ‘soft’ grain~\cite{BACHE2003,EVANS1994}. However, it has also been demonstrated that in some samples dwell sensitivity still occurs without 
the presence of a ‘rogue’ grain pair~\cite{BRANDES2010}, but the mechanism remains not well explained. The authors believe that understanding this behaviour of easy slip systems (basal and prism) during stress dwell will improve understanding of the complicated cold dwell fatigue problem.  

Although a considerable number of researchers have worked on this long-standing problem, a large proportion of the work has been performed at ambient temperatures. It has been suggested that ‘Cold dwell fatigue’ is strongly affected by temperature and dwell fatigue is suppressed by increasing temperature and disappears when temperature is above 200~$^{\circ}$C~\cite{TITANIUM}. Zhang \emph{et al}~\cite{ZHANG2015} investigated the temperature sensitivity of load shedding and cold dwell fatigue using the crystal plasticity model on a model Ti-6Al alloy. It was shown that at a temperature above 230~$^{\circ}$C thermally activated creep results in a rapid stress redistribution and elimination of local load shedding between ‘rogue’ grain pairs. An intermediate temperature of about 120~$^{\circ}$C exists at which the cold dwell debit is at its worst. However, the mechanism has not yet been verified by experimental observations. 

In this work, we combine a crystal plasticity finite element model and experimental methods to investigate the temperature effect on cold dwell fatigue in a coarse grain Ti-6Al alloy. An optical digital image correlation (DIC) approach has been used to allow in-situ measurement of strain in different grains while the samples were under load-up and load-hold (creep) type of deformation at 3 different temperatures up to 230~$^{\circ}$C. True sample microstructures, captured by electron back scattered diffraction (EBSD), were reconstructed for crystal plasticity simulations. Critical resolved shear stresses (CRSS) of the two main slip systems (basal and prism) were determined as a function of temperature by matching the experimentally measured strain to the simulated strain in different grains. Load shedding was studied in both ‘rogue’ and ‘non-rogue’ grain pairs at different temperatures. 

\section{Material and methods}
\subsection{Material}
\begin{figure}[hbt!]
\centering
\includegraphics[width=1\textwidth]{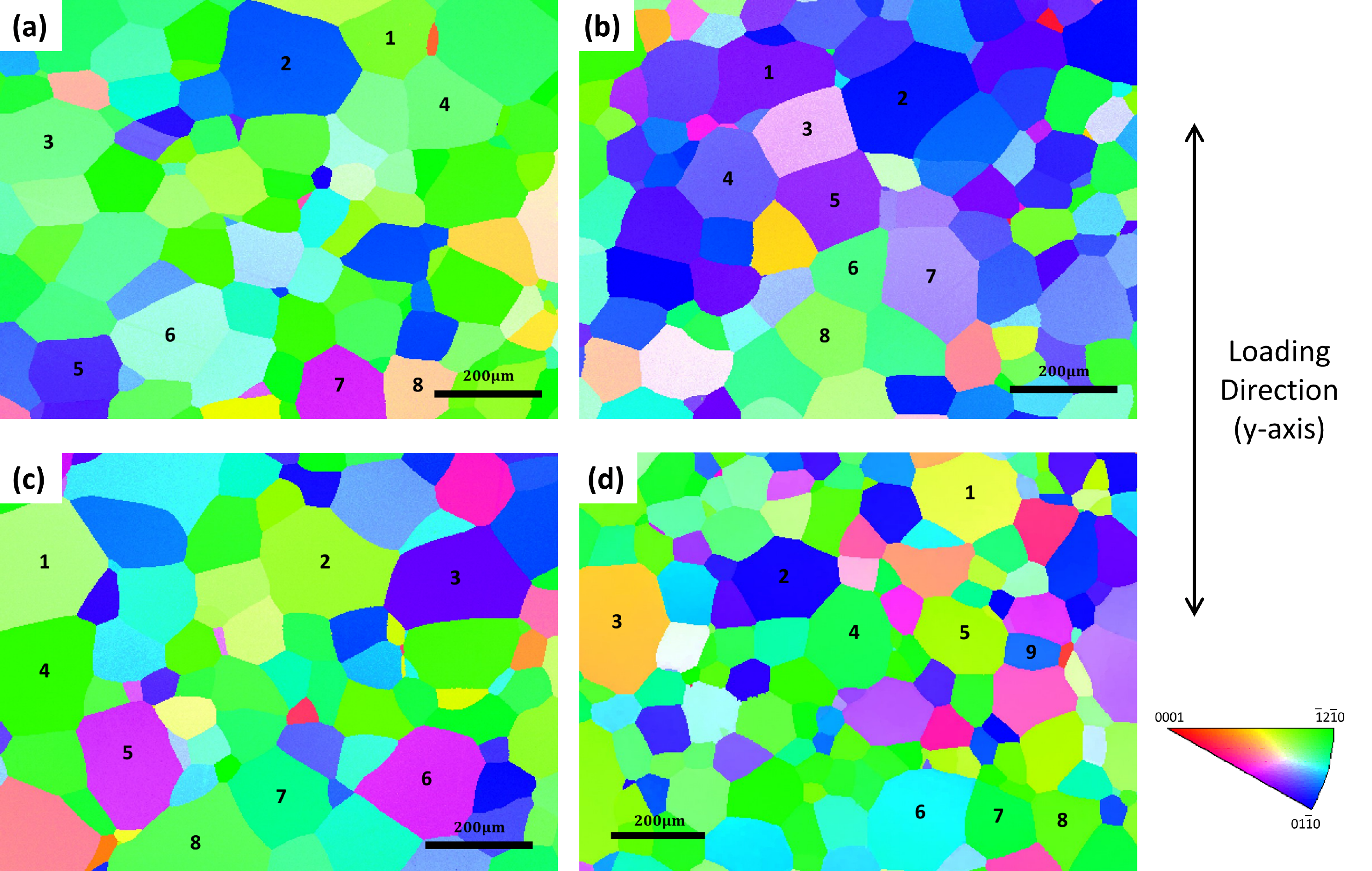}
\caption{EBSD maps of the Ti6Al samples (IPF colour map along the loading direction of the samples): (a) Sample 1 tested at room temperature; (b) Sample 2 tested at 120~$^{\circ}$C; (c) Sample 3 tested at 230~$^{\circ}$C and (d) Sample 4 tested at room temperature. Numbered grains 1-8 were selected to calibrate the CPFE model.}
\label{fig.1}
\end{figure}

The material used in this study is a Ti-6Al alloy bar, supplied by Timet UK Ltd. with composition shown in Table.\ref{table.1}. Test samples were  $5\times5\times5$ mm$^{3}$ cubes cut from the bar. In order to reduce the influence of pre-processing (e.g. residual stress and lattice defects) and homogenise the grain size, samples were annealed at 980~$^{\circ}$C in a vacuum tube furnace for 12~h, followed by slow cooling for a period of more than 24~h. The samples were metallographically prepared using SiC papers (up to 4000 grit), and a final polish with a $\approx$ 50~nm colloidal silica suspension. EBSD orientation maps were obtained using a Zeiss Merlin scanning electron microscope (SEM) equipped with a Bruker e-flash detector, operating at an accelerating voltage of 20~kV and a probe current of 20~nA. Figure~\ref{fig.1} shows the microstructure of these samples, the mean grain size after the heat treatment was found to be approximately 140$\pm$30~$\mu$m.

\begin{table*}[hbt!]
\centering
\caption{Composition of the Ti6Al alloy.}
 \begin{tabular*}{0.75\textwidth}{ @{\extracolsep{\fill}}c@{\extracolsep{\fill}}c@{\extracolsep{\fill}}c@{\extracolsep{\fill}}c@{\extracolsep{\fill}}c@{\extracolsep{\fill}}c@{\extracolsep{\fill}}} 
 \hline
 Ti & Al~(wt.~\%) &Fe~(wt.~\%) & O~(wt.~\%) & N~(wt.~\%) & Others total~(wt.~\%)\\
 \hline
 Balance & 5.79 &  $<$0.05 & 0.07 & 0.001 & 0.2 \\ 
 \hline
\end{tabular*}
\label{table.1}
\end{table*}

\subsection{Digital image correlation and creep tests}

\begin{figure}[hbt!]
\centering
\includegraphics[width=0.75\textwidth]{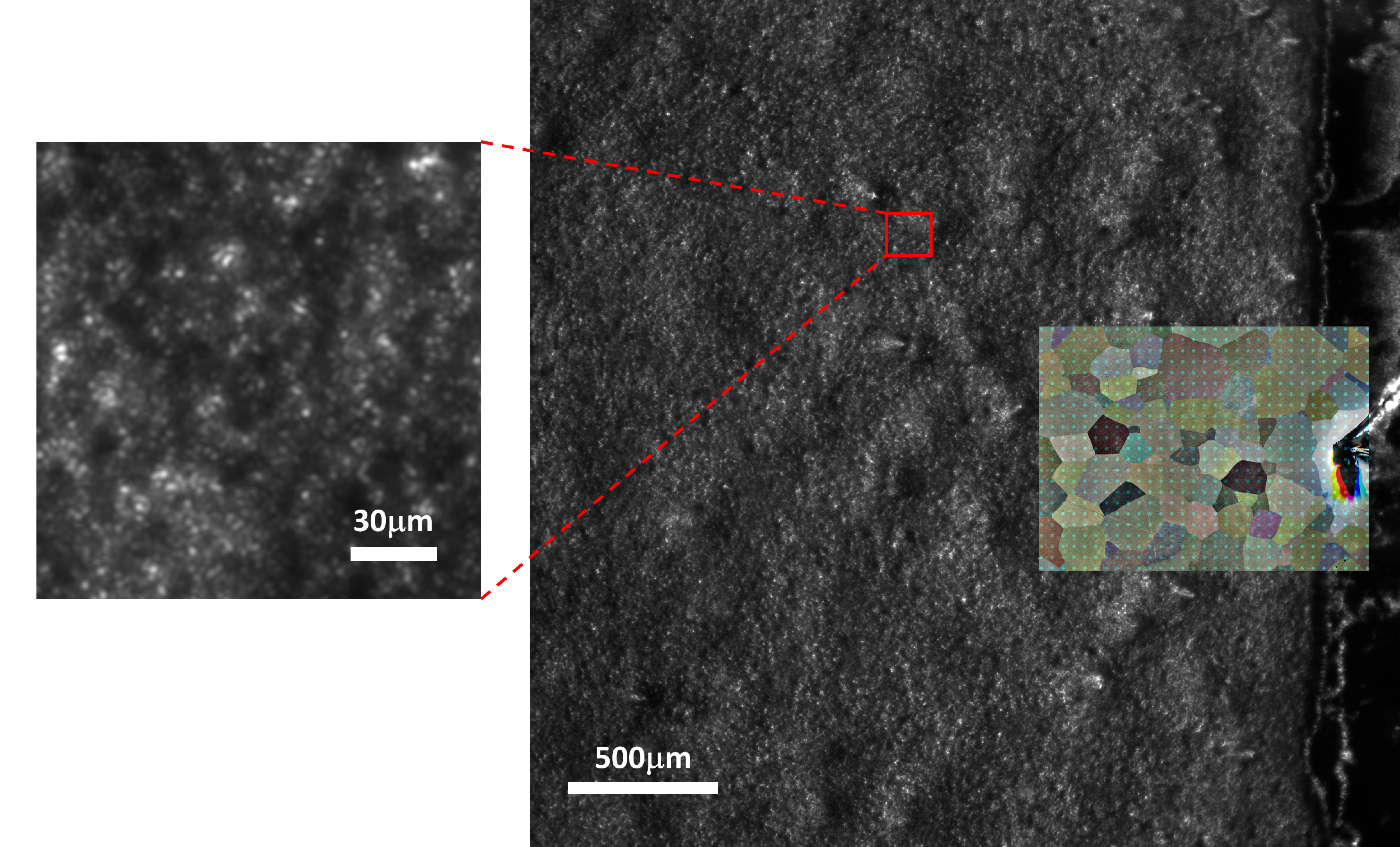}
\caption{Example image for DIC measurement, EBSD scanned area was located by matching a mark pre-scratched on the sample. Grid of nodes were generated within the EBSD scanned area. Loading along the vertical direction of this image.}
\label{fig.2}
\end{figure}

In order to measure strain on the sample surface, an optical DIC technique was used. This requires a speckle pattern to be applied to monitor pattern motion and subsequent strain estimation. This was achieved by spraying a polyurethane paint onto the polished surfaces of the cubic samples. Black paint was used first to generate a uniform background, then a white paint diluted with isopropyl alcohol was applied to the surfaces using an airbrush kit to generate speckle patterns. An example of the speckle pattern is shown in Figure~\ref{fig.2}, the size of the white spots vary from 1-10~$\mu$m. The images were recorded at 16-bit depth using a high speed camera (Hamamatsu Photonics Ltd - ORCA-Flash 4.0) with an image resolution of 2048$\times$2048 pixels. A Questar QM100 lens attachment was employed to enable imaging the sample through a furnace window for elevated temperature experiments. With the working distances of $\approx$0.5~m used in this experiment, the pixel size corresponds to 1.5~$\mu$m.

\begin{figure}[hbt!]
\centering
\includegraphics[width=0.55\textwidth]{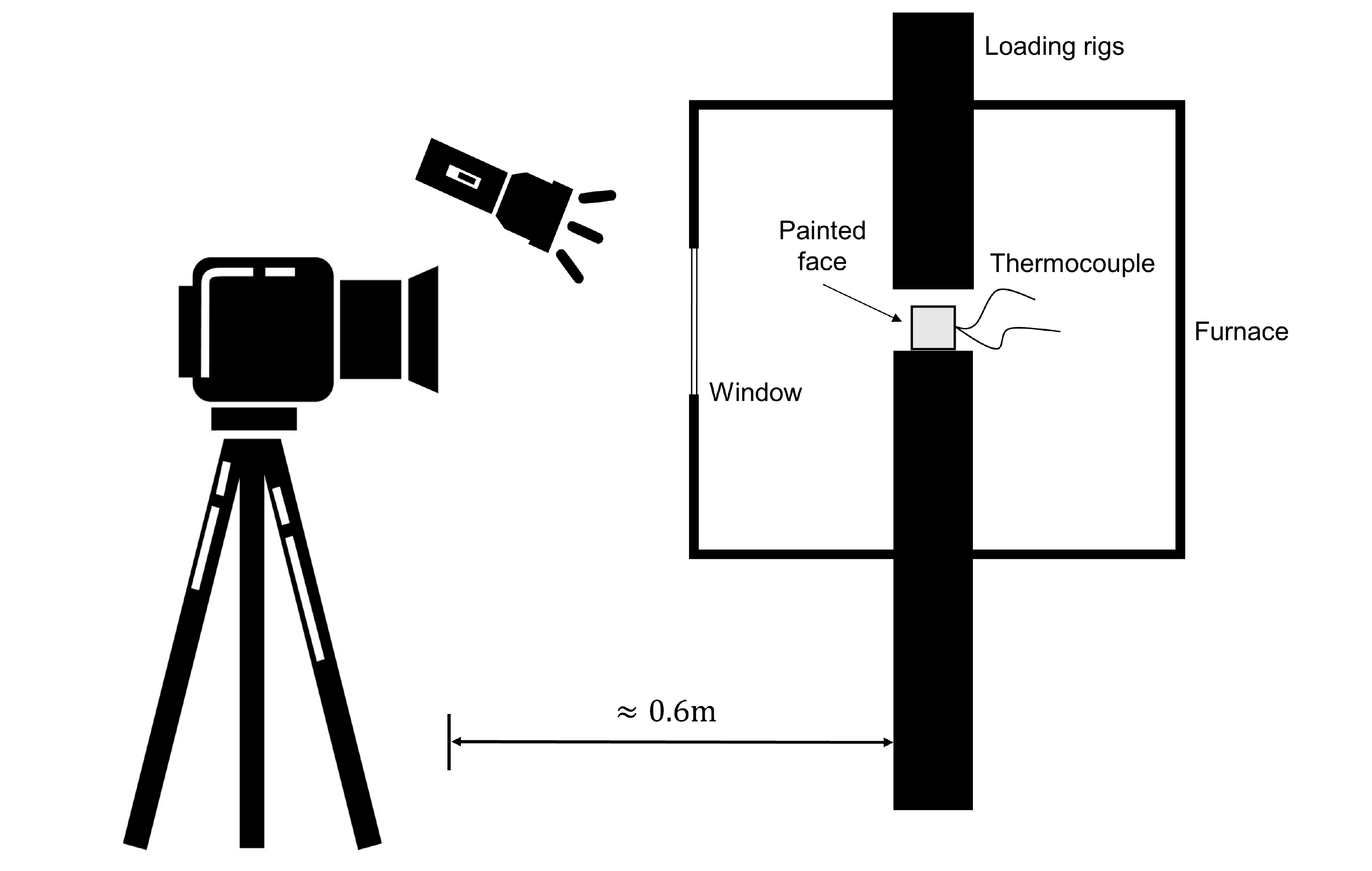}
\caption{Sketch of the experiment setup.}
\label{fig.3}
\end{figure}

Figure~\ref{fig.3} shows the experiment setup. Samples with thermocouples spot welded on one face, were placed in an environmental chamber (manufactured by Severn Thermal Solutions). DIC images were recorded by the high speed camera through the furnace window. An LED light source was placed outside the furnace chamber, illuminating the sample surfaces with speckle patterns, so that the exposure time of the camera could be reduced down to 20~ms without sacrifice the quality of the images. The acquisition rate was set to be constant at 1 frame per second for all the tests. Sample 1 to sample 3 were tested at room temperature, 120~$^{\circ}$C and 230~$^{\circ}$C (ambient temperature, intermediate temperature and temperature where cold dwell fatigue diminished). Compressive loads were applied to the samples using a Shimadzu mechanical test frame (AGS-X series) after the temperature of each sample was stabilised at each target temperature for 5 minutes.    

\begin{figure}[hbt!]
\centering
\includegraphics[width=1\textwidth]{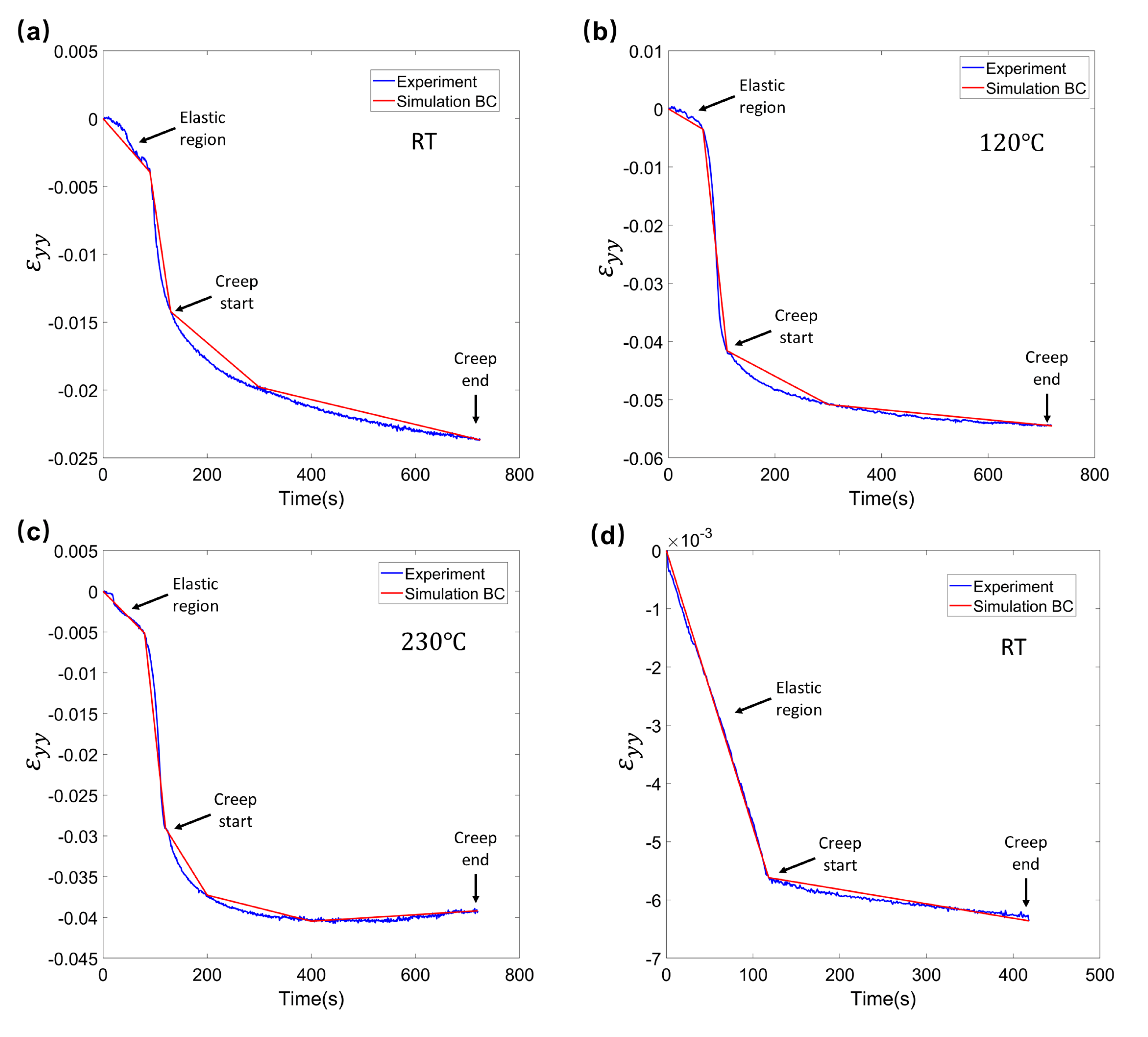}
\caption{Strain vs. time curves of (a) Sample 1 loaded beyond macroscopic yield stress at room temperature; (b) Sample 2 loaded beyond macroscopic yield stress at 120~$^{\circ}$C; (c) Sample 3 loaded beyond macroscopic yield stress at 230~$^{\circ}$C and (d) Sample 4 loaded below macroscopic yield stress at room temperature.}
\label{fig.4}
\end{figure}

Samples were loaded in compression at constant stress rate of 5~MPa/s, so that the first gradient changes observed in the strain vs. time plots (see Figure~\ref{fig.4}(a)-(c)) represent the macroscopic yield. Sample 1-3 were designed to study the behaviour of the two major slip systems (basal and prism slip) during creep and thus were loaded beyond the macroscopic yield stresses to activate both the desired slip systems. After clear gradient changes were observed in strain vs. time data, the macroscopic stresses were held constant for a period of 10 minutes to mimic the dwell effect. The compression strain of these samples was found to increase during the stress dwell periods. In the first half of this dwell period, strain appears to change faster than that of the subsequent half. This also shows that stress redistribution is more active in the first 5 minutes of the stress hold. Sample 4 was designed to scope into this first 5 minutes of creep under a stress below the macroscopic yield stress, which is a better representation of a real load condition of Ti alloys in-service. The loading curve of sample 4 is shown in Figure~\ref{fig.4}(d) and this sample starts to creep with strain still in elastic region, and lasts for 5 minutes.       

As shown in Figure~\ref{fig.2}, identification marks were scratched on the painted surfaces so that the area of the EBSD scan can be correlated to the DIC strain measurements. A grid of nodes were generated within the EBSD scanned area, with intervals of 25 pixels in the optical images ($\approx$37.5~$\mu$m). DIC strain measurements were performed on the images captured by the camera via a modified version of the Python based DIC code~\cite{PYDIC} so that the in-plane strain components at the nodes can be calculated. The $\varepsilon_{yy}$ components of the nodes in the central regions of the selected grains (dismissing the nodes near the grain boundaries) were averaged to represent the averaged strain along the load direction within particular grains. The selected grains have a relatively larger area and contain more nodes and thus the average values have a higher confidence when correlating areas between EBSD and DIC. The strains within these selected grains were then used to calibrate the CPFE model.

\subsection{Crystal plasticity model and simulations}
\label{sec:CPmodel}

A crystal plasticity hypoelastic formulation was used as a constitutive model for the finite element simulations~\cite{DUNNE20071061}. It is implemented as a UMAT for Abaqus based on the user-defined elements (UEL) subroutine by Dunne \textit{et al}~\cite{DUNNE20071061} and it is based on the finite strain theory. The temperature dependent elastic constants of the Ti-6Al alloy are shown in Table~\ref{table.2}. The constitutive model calculates the Cauchy stress $\boldsymbol{\sigma}$ based on the total strain increment $\Delta \boldsymbol{\varepsilon}$ at each time step, which is calculated from the deformation gradient $\mathbf{F}$ provided by the finite element solver. The plastic strain $\boldsymbol{\varepsilon}_{p}$ is the internal variable that quantifies the irreversible plastic deformation induced by the slip movement of dislocations. The time evolution of the plastic strain takes into account $N_{\textrm{slip}}=$30 slip systems in the HCP crystal \cite{KALIDINDI1998267,Grilli2020PRM}:

\begin{equation}
\dot{\bm{\varepsilon}}_\textrm{p} \left ( \boldsymbol{\sigma} \right ) =  \frac{1}{2} \sum_{\alpha=1}^{N_{\textrm{slip}}} \dot{\gamma}_{\alpha} \left ( \bm{\sigma} \right ) \left ( \bm{s}_\alpha \otimes \bm{n}_\alpha + \bm{n}_\alpha \otimes \bm{s}_\alpha \nonumber \right ) \ ,
\label{eqn:plasticvelocitygrad}
\end{equation}
where $\dot{\gamma}_{\alpha}$ is the plastic strain rate, $\bm{s}_\alpha$ the slip direction and $\bm{n}_\alpha$ the slip plane normal of the slip system $\alpha$. $\bm{s}_\alpha$ and $\bm{n}_\alpha$ are expressed in the sample reference frame. 


%
\begin{table*}[!htb]
    \centering
    \caption{Elastic constants (GPa) of Ti6Al in Voigt notation \cite{DUNNE20071061,ZHANG2015}.}
    \begin{tabular}{|c|c|c|c|c|c|c|c|c|}
         \hline
         Temperature & $\mathbb{C}_{11}$,$\mathbb{C}_{22}$ & $\mathbb{C}_{33}$ & $\mathbb{C}_{12}$ & $\mathbb{C}_{13}$,$\mathbb{C}_{23}$ & $\mathbb{C}_{44}$,$\mathbb{C}_{55}$ & $\mathbb{C}_{66}$ \\
         \hline
         20~$^{\circ}$C & 139.21 & 162.57 & 81.95 & 68.78 & 42.50 & 30.02 \\
         120~$^{\circ}$C & 126.48 & 147.63 & 74.45 & 62.46 & 36.10 & 26.20 \\
         230~$^{\circ}$C & 112.52 & 131.35 & 66.23 & 55.57 & 31.90 & 23.20 \\
         \hline
    \end{tabular}
    \label{table.2}
\end{table*}

The slip systems used are reported in Table \ref{tab:sliptwinsystems}. $\bm{s}_\alpha^0$ and $\bm{n}_\alpha^0$ are the slip directions and normals in the crystal reference frame.

\begin{table*}[htb]
    \centering
    \caption{Slip systems used in the model. Slip directions and normals are expressed in the crystal reference frame. The vectors are normalised.}
    \begin{tabular}{|l|c|c|}
         \hline
         Slip system & $\bm{s}_{\alpha}^0$ & $\bm{n}_{\alpha}^0$ \\
         \hline
         $<$a$>$ basal slip & $\left [ -0.5, \pm 0.866,  0 \right ]$ & $\left [ 0, 0, 1 \right ]$ \\
          & $\left [ 1 , 0 , 0 \right ]$ & $\left [ 0 , 0 , 1 \right ]$ \\
         $<$a$>$ prismatic slip & $\left [ -0.5, \pm 0.866,  0 \right ]$ & $\left [ \mp 0.866, -0.5, 0 \right ]$ \\
          & $\left [ 1, 0, 0 \right ]$ & $\left [ 0, -1, 0 \right ]$ \\
         $<$a$>$ 1\textsuperscript{st} order pyramidal slip & $\left [ -1, 0, 0 \right ]$ & $\left [ 0, \mp 0.7559, 0.6547 \right ]$ \\
         & $\left [ -0.5, -0.866, 0 \right ]$ & $\left [ \pm 0.6547, \mp 0.3780, 0.6547 \right ]$ \\
         & $\left [ -0.5, 0.866, 0 \right ]$ & $\left [ \mp 0.6547, \mp 0.3780, 0.6547 \right ]$ \\
         $<$c+a$>$ 1\textsuperscript{st} order pyramidal slip & $\left [ -0.3536, 0.6124, \pm 0.7071 \right ]$ & $\left [ \pm 0.6547, \mp 0.3780, 0.6547 \right ]$ \\
          & $\left [ -0.3536, -0.6124, \mp 0.7071 \right ]$ & $\left [ -0.6547, -0.3780, \pm 0.6547 \right ]$ \\
          & $\left [ 0.7071, 0, \pm 0.7071 \right ]$ & $\left [ \mp 0.6547, \pm 0.3780,  0.6547 \right ]$ \\
          & $\left [ \mp 0.7071, 0, 0.7071 \right ]$ & $\left [ \pm 0.6547, \pm 0.3780,  0.6547 \right ]$ \\
          & $\left [ -0.3536, -0.6124, \pm 0.7071 \right ]$ & $\left [ 0, \pm 0.7559, 0.6547 \right ]$ \\
          & $\left [ \pm 0.3536, \mp 0.6124, 0.7071 \right ]$ & $\left [ 0, \pm 0.7559, 0.6547 \right ]$ \\
         $<$c+a$>$ 2\textsuperscript{nd} order pyramidal slip & $\left [ -0.3536, 0.6124, \pm 0.7071 \right ]$ & $\left [ \pm 0.3536, \mp 0.6124,  0.7071 \right ]$ \\
          & $\left [ -0.3536, -0.6124,  \pm 0.7071 \right ]$ & $\left [ \pm 0.3536, \pm 0.6124,  0.7071 \right ]$ \\
          & $\left [ 0.7071, 0,  \pm 0.7071 \right ]$ & $\left [ \mp 0.7071,  0.0000,  0.7071 \right ]$ \\
         \hline
    \end{tabular}
    \label{tab:sliptwinsystems}
\end{table*}

The plastic strain rate on each slip system is calculated using a physics-based law (Orowan's equation) that is a function of the resolved shear stress (RSS) $\tau_\alpha = \bm{s}_\alpha \cdot \bm{\sigma} \cdot \bm{n}_\alpha$ on the respective slip system \cite{DUNNE20071061,Orowan1934,DAS201818,ROTERS2019420}:

\begin{equation}
\dot{\gamma}_\alpha \left ( \bm{\sigma} \right ) = \rho_m b_\alpha^2 \nu \exp \left ( - \frac{\Delta F_\alpha}{kT} \right ) \sinh \left ( \frac{ \left (\left | \tau_\alpha \right | - \tau_\alpha^c \right ) \Delta{V}_\alpha }{k T}  \right ) \mathrm{sign}(\tau_\alpha) H \left ( \left | \tau_\alpha \right | - \tau_\alpha^c \right ) \ .
\label{eqn:sliprateforwardbackward}
\end{equation}
$\rho_m$ is a constant, representing the average value of the mobile dislocation density. $b_\alpha$ is the Burgers vector magnitude and $\nu$ is the characteristic frequency at which a dislocation attempts to overcome the Helmoltz free energy barrier $\Delta F_\alpha$. Once this barrier is overcome, the dislocation movement can induce plastic slip. $\Delta{V}_\alpha$ is the activation volume of the process. $k$ is the Boltzmann constant and $T$ is the temperature. $\tau_\alpha^c$ is the critical resolved shear stress (CRSS) of the slip system $\alpha$ and $H$ is the Heaviside function. Therefore, plastic deformation on one slip system is induced only when the magnitude of RSS exceeds the CRSS. 

At each time increment $\Delta t$, the increment of the Cauchy stress in the sample reference frame is calculated as \cite{HILL1972401,GRILLI2020109276}:

\begin{align}
\label{eqn:cauchystressincr}
\Delta \boldsymbol{\sigma} &= \mathbb{C} \Delta \boldsymbol{\varepsilon}_{e} + \left ( \bm{W}_e \boldsymbol{\sigma}_0 - \boldsymbol{\sigma}_0 \bm{W}_e \right ) \Delta t \\
&= \mathbb{C} \left ( \Delta \boldsymbol{\varepsilon} - \dot{\bm{\varepsilon}}_\textrm{p} \left ( \boldsymbol{\sigma} \right ) \Delta t \right ) + \left ( \bm{W}_e \boldsymbol{\sigma}_0 - \boldsymbol{\sigma}_0 \bm{W}_e \right ) \Delta t \ , \nonumber
\end{align}
where $\boldsymbol{\sigma}_0$ is the Cauchy stress at the previous increment and $\bm{W}_e$ is the elastic continuum spin \cite{Belytschko2014}. Reorientation of the grains due to deformation is also included in the model.
The constant parameters input into the model are summarised in Table.~\ref{tab:modelparameters}. 
\begin{table*}[htb]
\centering
\caption{Constant parameters in the model~\cite{DUNNE20071061,ZHANG2016393,ZHENG2016411}.}
\def\arraystretch{1.2}
\begin{tabular}{|l|c|c|}
\hline
Mobile dislocation density & $\rho_m$ & 5 $\mu$m$^{-2}$ \\
\hline
Burgers' vector (basal and prismatic slip) & $b_\alpha$ & 0.295 nm \\
\hline
Burgers' vector (pyramidal slip) & $b_\alpha$ & 0.5533 nm \\
\hline
Characteristic attempt frequency & $\nu$ & $1\times10^{11}$ s$^{-1}$ \\
\hline
Boltzmann constant & $k$ & 1.38$\times10^{-23}$~JK$^{-1}$\\
\hline
\end{tabular}
\label{tab:modelparameters}
\end{table*}

The other three terms, $\Delta F_\alpha$, $\Delta{V}_\alpha$ and $\tau_\alpha^c$ are temperature dependent. Values of $\Delta F_\alpha$ and $\Delta{V}_\alpha$ input into the model can be found in Figure~\ref{fig.6}, following the previous work~\cite{XIONG2020_2}. $\tau_\alpha^c$ for prismatic and basal slip were the target values we seek to obtain in the optimisation procedure carried out in this work. $<c+a>$ Pyramidal slip is much more difficult to activate due to its higher $\tau_\alpha^c$, which is about three times the value of basal slip at temperature between 20~$^{\circ}$C and 300~$^{\circ}$C~\cite{HASIJA2003,ZHANG2015,Williams2002}. The values of $\tau_\alpha^c$ for pyramidal slip was set to be three time the value of basal slip in this work.    

The geometrically necessary dislocation (GND) density is calculated using a least square minimisation~\cite{ARSENLIS19991597}. The sum of the squares of the GND densities on the different slip systems: $\sum_{\alpha=1}^{N_{\textrm{slip}}} \rho_\alpha^2$ is minimised. A constrained minimisation algorithm is applied. Given the simulated plastic deformation gradient $\mathbf{F}_p$, the GND densities $\rho_\alpha$ must satisfy the condition:

\begin{equation}
\left ( \nabla \times \mathbf{F}_p \right )^T = \sum_{\alpha=1}^{N_{\textrm{slip}}} \rho_\alpha \bm{b}_\alpha \otimes \bm{l}_\alpha \ ,
\label{eqn:nyetensor}
\end{equation}
where $\bm{b}_\alpha$ is the Burgers vector and $\bm{l}_\alpha$ is the dislocation line unit vector for the $\alpha^{th}$ slip system. The right-hand side of equation (\ref{eqn:nyetensor}) is known as Kr\"oner-Nye tensor \cite{NYE1953153}. The dislocation energy is minimised while respecting the condition in equation (\ref{eqn:nyetensor}), therefore the different $\rho_\alpha$ can be calculated at each time step of the simulation. More details of the algorithm are given in \cite{DAS201818}. Analysing the GND density is useful to understand the relative activation of the slip systems and the incompatibility of the plastic deformation between neighbouring grains, as described in section \ref{sec:sliptrace}. The crystal plasticity code is available in the following repository \cite{CrystalPlasticityUMAT}.

\begin{figure}[hbt!]
\centering
\includegraphics[width=0.75\textwidth]{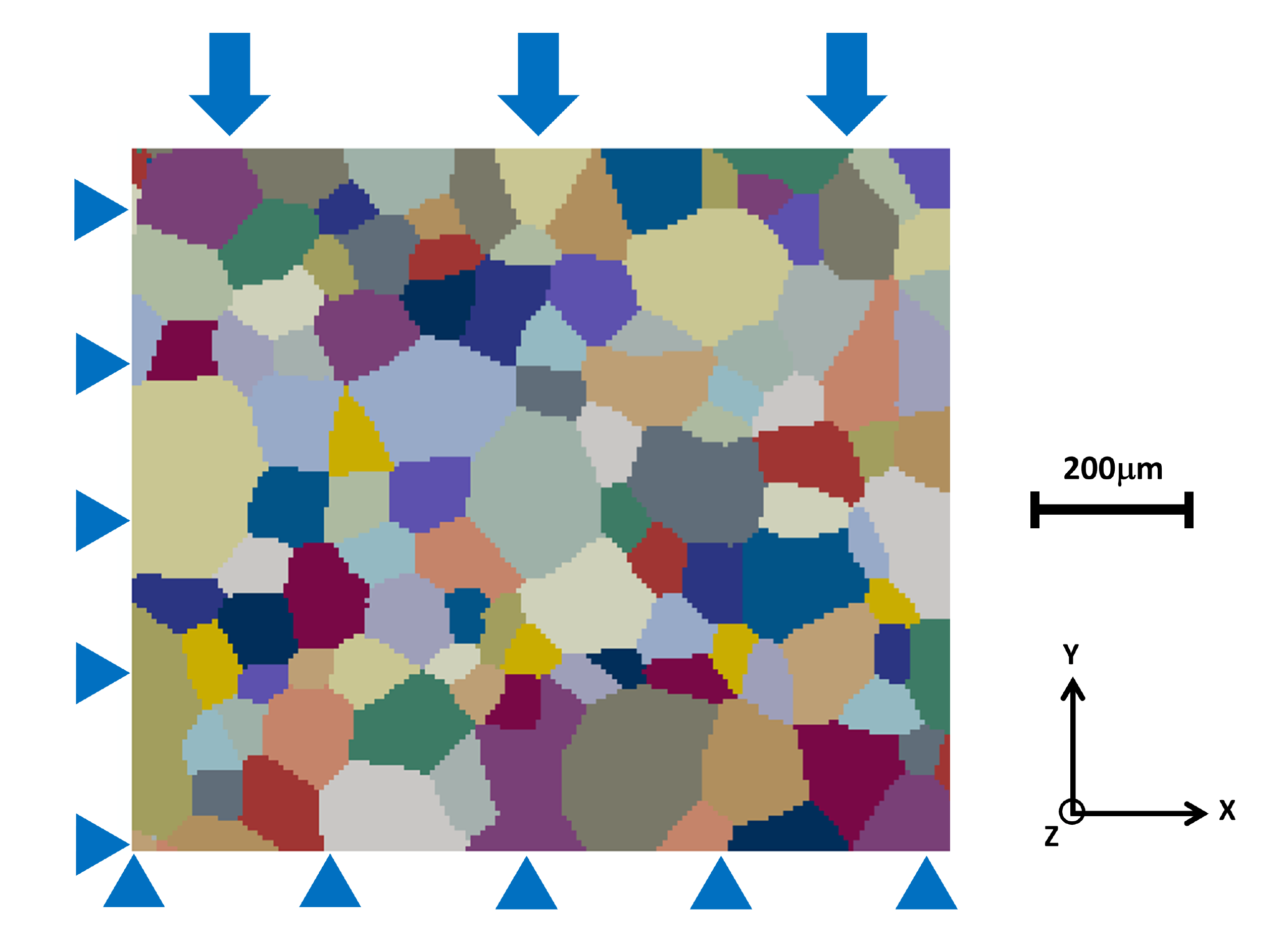}
\caption{Example of the meshed model of sample 4, the depth of the model was 2 elements. Arrows show the boundary condition of the model. Grains were rendered with different colours so that the microstructure was found to consistent with the EBSD map.}
\label{fig.5}
\end{figure}

\begin{figure}[hbt!]
\centering
\includegraphics[width=1\textwidth]{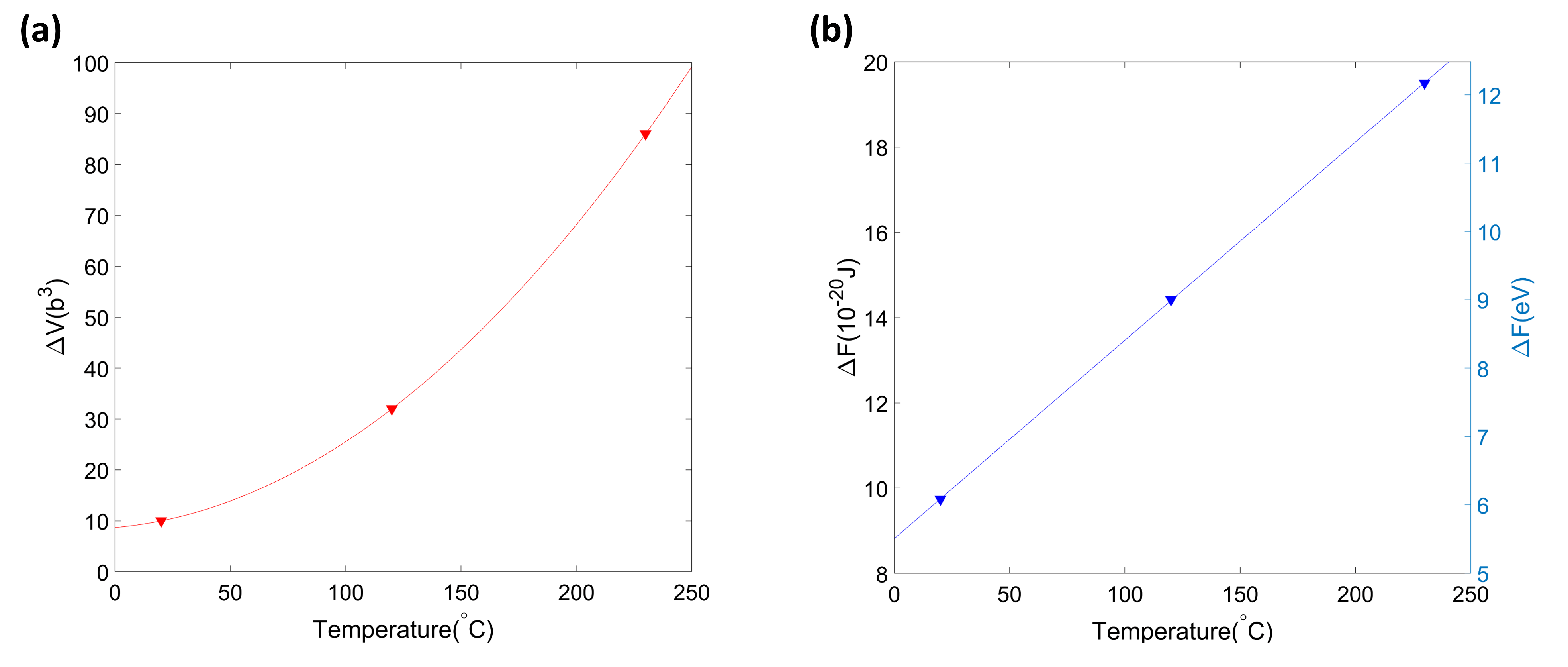}
\caption{Input parameters of (a) activation volume ($\Delta{V}$) and (b) activation energy ($\Delta{F}$) as a function of temperature~\cite{XIONG2020_2}.}
\label{fig.6}
\end{figure}
 
Figure~\ref{fig.5} shows the representative volume of sample 4, which is generated using the EBSD map and a MATLAB function called ebsd2abaqus~\cite{EBSD2ABAQUS}. Grain morphology and orientations were the same as the real samples. The element type was set to be quadratic with reduced integration (C3D20R) and meshed as a regular, equisized grid of size matched to 8~$\mu$m. The model was a 3D model with a depth of 2 elements, the 2 layers of elements were identical (validation can be found in the Appendix). The movement of the left face, bottom face and back face were constrained along the X, Y and Z direction respectively. Displacement was applied to the top face so that the overall strain of the model matches with the averaged strain of the sample (as shown in Figure~\ref{fig.4}). Within each grain, nodes on the front surface in the central region of the grain (again the nodes near the grain boundaries were dismissed, to be a closer estimate to DIC measurements) were grouped to represent the grain. Strain component $\varepsilon_{yy}$ of these nodes was averaged and matched with the experimentally measured DIC strain, as described in the next section.   

\section{Results}
\subsection{Critical resolved shear stress}

The CRSS for the two main slip systems (prism and basal) were obtained by calibrating the CPFE model. The optimisation was achieved using a Scipy package~\cite{SciPy}, by use of the Nelder-Mead method~\cite{NelderMead1965,GRILLI2020103800}. In a typical loop, a set of initial guesses of the CRSS values were input into the model and the simulation job was submitted. Once the simulation was complete, strains from the 8 selected grains in each sample were compared with the experimental strain results of these grains. The differences between simulated strains and experimental strains were calculated and was aimed to be minimised. The Nelder-Mead algorithm then generated a new set of CRSS to re-run the simulation. This iterative process was carried out until the difference between simulated and experimental strain in the grains was minimal. 

Figure~\ref{fig.7} shows the comparison of the strains within the 8 chosen grains between experiments and simulations. The optimisation method was applied to samples 1-3, after which the difference between simulated and experimental strain in the grains becomes lower than 10\%. All 3 samples show good agreement in experimental and simulated strains. The CRSS values of prismatic and basal slip systems were extracted at these three temperatures (as shown in Figure~\ref{fig.8}(a)). CRSS values at room temperature were then used for sample 4. As shown in Figure~\ref{fig.7}(g) and Figure~\ref{fig.7}(h), the simulated strains still fit well with the experimental strains. 

The relationship between $\tau_\alpha^c$ and temperature can be fitted using an empirical power law function~\cite{KISHIDA2020168,MITCHELL20019827}. The resulting temperature dependence of the CRSS of the prismatic and basal slip systems are given by: 
\begin{equation}
\tau_{basal}^c=1054.6{T}^{-0.472}\ \textrm{(MPa)}
\end{equation}
\begin{equation}
\tau_{prism}^c=733.8{T}^{-0.4622} \ \textrm{(MPa)}
\end{equation}
where $T$ is the temperature in $^{\circ}$C.

There are also other workers fitted the CRSS vs. temperature relationship with quadratic functions~\cite{ZHANG2015,ZHENG2018}. However, for our data-set, false minima were found at relatively low temperatures (150-200~$^{\circ}$C as shown in Figure~\ref{fig.8}(b)). Although quadratic function can provide a better fitting, the trend of CRSS vs. temperature predicted by quadratic functions cannot reflect the true material behaviour. 

\subsection{Slip activity and trace analysis}
\label{sec:sliptrace}

Figure~\ref{fig.9} shows the geometrically necessary dislocations (GND) density of the three prism and the three basal slip systems in sample 4 at the end of creep. These maps were meant to identify the active slip systems within different grains. At the stress level of sample 4, the basal slip was barely activated. By contrast, prism slip was activated in some specific grains, as shown by the Prism \#1 and Prism \#3 GND density in Figure~\ref{fig.9}(d) and (f). There are two regions of interests: pair 1 (grain 5 \& 9) represents the ‘rogue grain pair’ where only one grain (grain 9) activates plastic slip; pair 2 (grain 6 \& 7) represents the ‘non-rogue grain pair’ where both grains activate plastic slip. 

The simulated slip systems activities matches well with the experimental observations. Figure~\ref{fig.10}
shows the comparison of grain morphology of the two grain pairs before and after the experiment. Slip traces were observed in grain 6, grain 7 and grain 9 but not in grain 5. All the newly generated slip lines were found to be parallel within each grain, indicating only one slip system was activated for each grain. This agreed well with the simulated slip activity, where Prism 1 was activated in grain 7, Prism 3 was activated in grain 6 and grain 9. Slip lines simulated by the CPFE model were plotted for better comparison with the experimental observations. To do this, we used the method introduced by Guery \textit{et al}~\cite{GUERY2016}, where the slip plane normal, $\textbf{\textit{n}}_\alpha$ of the active slip system $\alpha$ was rotated into the sample coordinate system by the grain rotation matrix, \textit{\textbf{R}}. The slip line vector, $\textbf{\textit{t}}_\alpha$ was obtained by the cross product of the rotated slip plane normal $\textbf{\textit{n}}_\alpha^{s}$ and the sample surface normal $\textit{\textbf{e}}_{z}$ as:
\begin{equation}
    \textbf{\textit{t}}_\alpha = \textbf{\textit{n}}_\alpha^{s} \times \textit{\textbf{e}}_{z}
\end{equation}

These line segments from every element within grain 6, grain 7 and grain 9 were then plotted (as shown in Figure~\ref{fig.10}). There is good agreement between alignment of slip traces calculation in the model and experimental observation. The length of the segment was proportional to the magnitude of the averaged GND density of each element. The ratio of the averaged GND density in grain 6, grain 7 and grain 9 was approximately 8:10:5 (i.e. the slip activity was the highest in grain 7 and the lowest in grain 9). This is also consistent with the slip line density observed in these grains.      

\subsection{Load shedding}

As shown in Figure~\ref{fig.11}(a) and \ref{fig.11}(b), along path A and path B in grain pair 1 and 2 respectively, the load shedding effect in ‘rogue’ and ‘non-rogue’ grain pairs has been studied. Figure~\ref{fig.11}(c)-(h) show the stress evolution along the paths at 3 different temperatures. At room temperature, load shedding was obvious in the ‘rogue’ grain pair (pair 1). Compared to the start of the creep loading, the stress at the end of creep within the ‘hard’ grain was found to increase while the stress within the ‘soft’ grain remained approximately at the same level. As a result, the stress difference near the grain boundary increased to a higher level of approximately 100~MPa. In comparison, load shedding was not as significant in the ‘non-rogue’ grain pair, where stress changes in both grains before and after the creep are smaller and the stress difference at the grain boundary after creep was only approximately 50~MPa. 

Higher temperature simulations have been performed on the model of sample 4 for better comparison, the parameters (CRSS, $\Delta{F}$, $\Delta{V}$ and elastic constants) were changed to parameters at 120~$^{\circ}$C and 230~$^{\circ}$C but the boundary condition remained the same. At higher temperatures, the stresses in all these four grains were found to relax after creep, indicating plasticity. This is also confirmed by the increase of GND density in these grains as temperature rises and shown in Figure~\ref{fig.11}(i) and \ref{fig.11}(j). GND densities at the grain boundaries were significantly higher than that within the grains, showing a dislocation pile-up~\cite{XU2020,Stroh1954} at the grain boundaries. For the rogue grain pair, the initial stress difference across the grain boundary increases slightly for $T=120^{\circ}$C compared to room temperature, then drops at higher temperature. But during the load hold, the stress difference does not increase as some plasticity is now also possible in the hard grain (non-zero GND density in the hard grain in Figure~\ref{fig.11}(i)).  

\begin{figure}[H]
\centering
\includegraphics[width=0.9\textwidth]{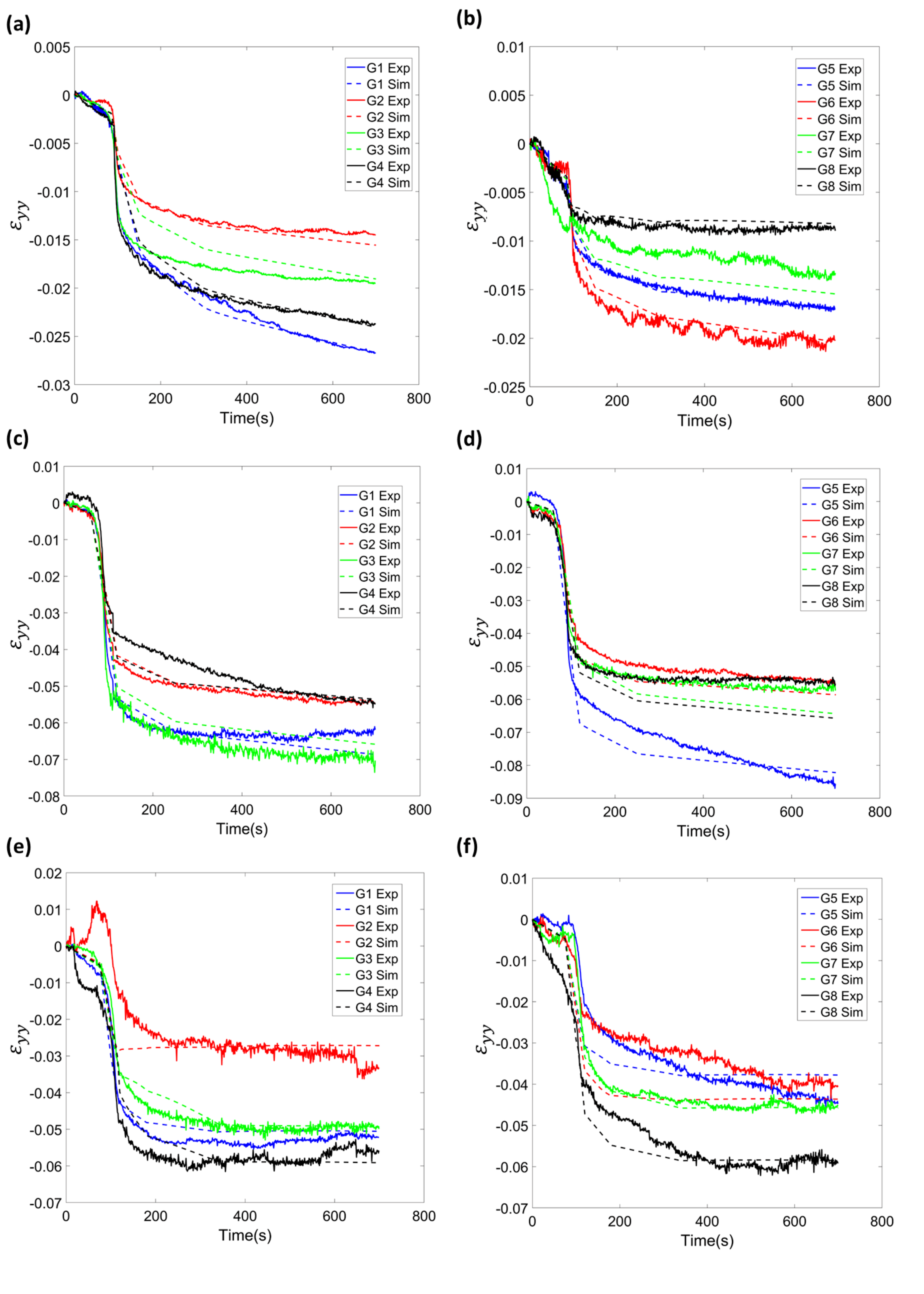}
\phantomcaption
\end{figure}
\begin{figure}[hbt!]\ContinuedFloat
\centering
\includegraphics[width=0.9\textwidth]{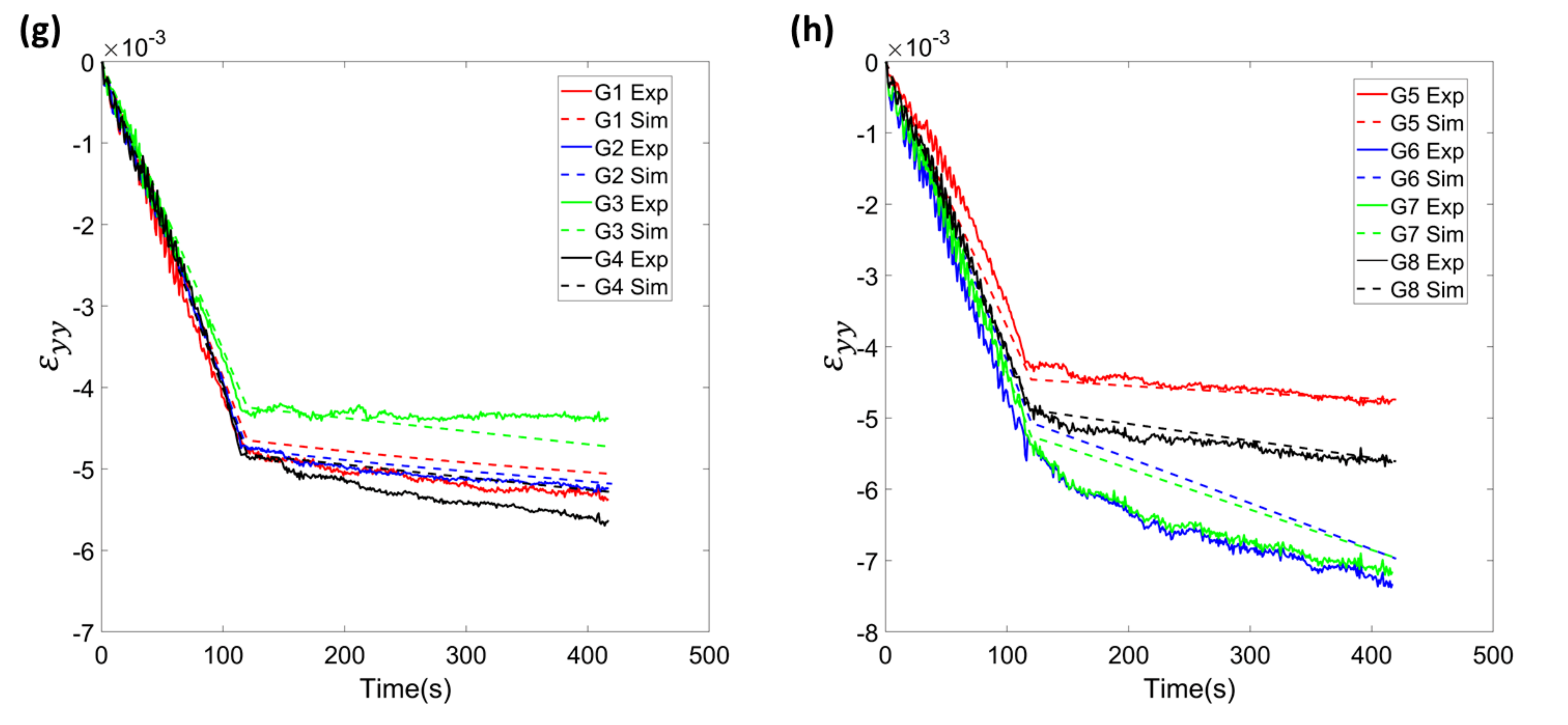}
\caption{Strain vs. time plots of experiments and simulations of (a) sample 1 grain 1-4; (b) sample 1 grain 5-8; (c) sample 2 grain 1-4; (d) sample 2 grain 5-8; (e) sample 3 grain 1-4; (f) sample 3 grain 5-8; (g) sample 4 grain 1-4; (h) sample 4 grain 5-8 (Note that the scales of the vertical axes are different the plots).}
\label{fig.7}
\end{figure}

\begin{figure}[hbt!]
\centering
\includegraphics[width=0.95\textwidth]{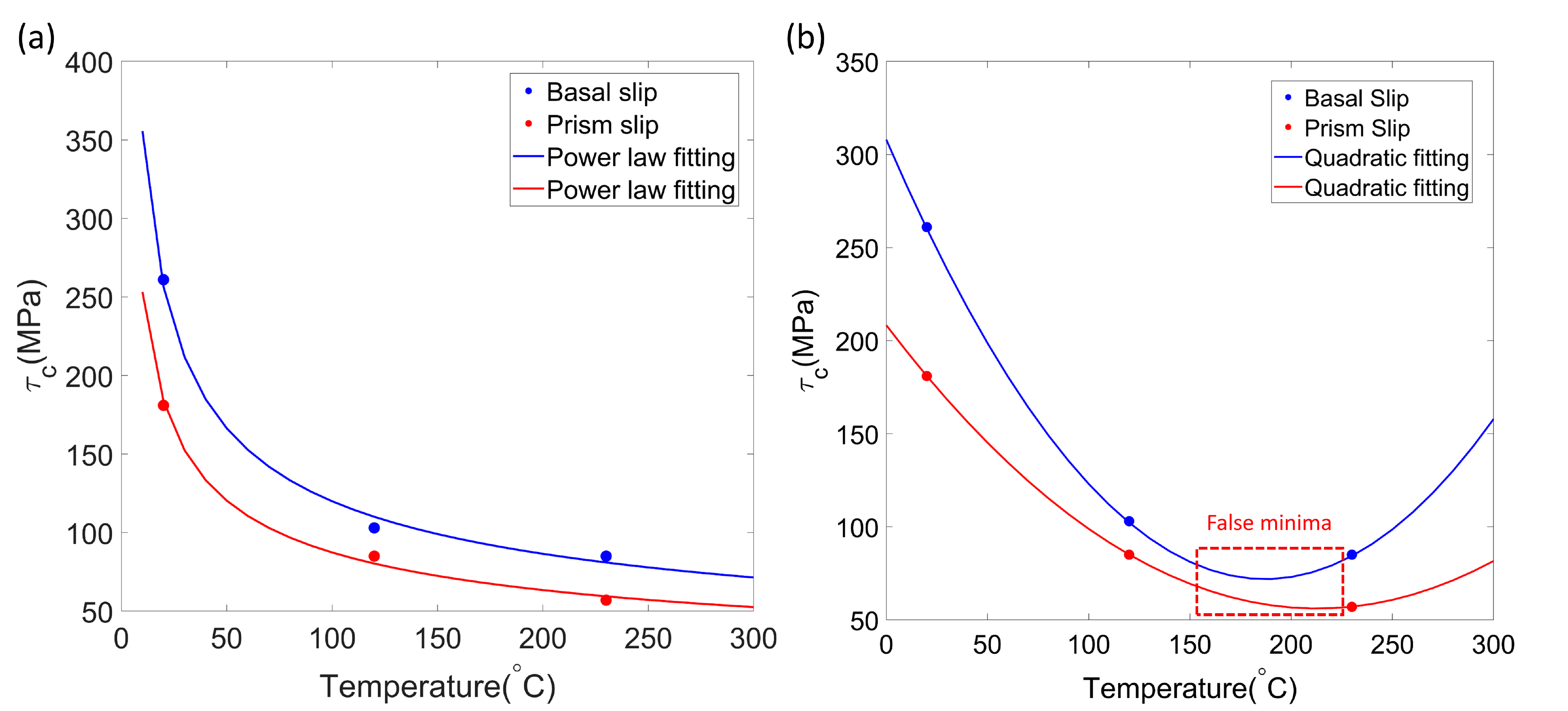}
\caption{Calibrated CRSS values of prismatic and basal slip systems as a function of temperature: (a) fitted with power law functions; (b) fitted with quadratic functions. }
\label{fig.8}
\end{figure}

\clearpage

\begin{figure}[H]
\centering
\includegraphics[width=1\textwidth]{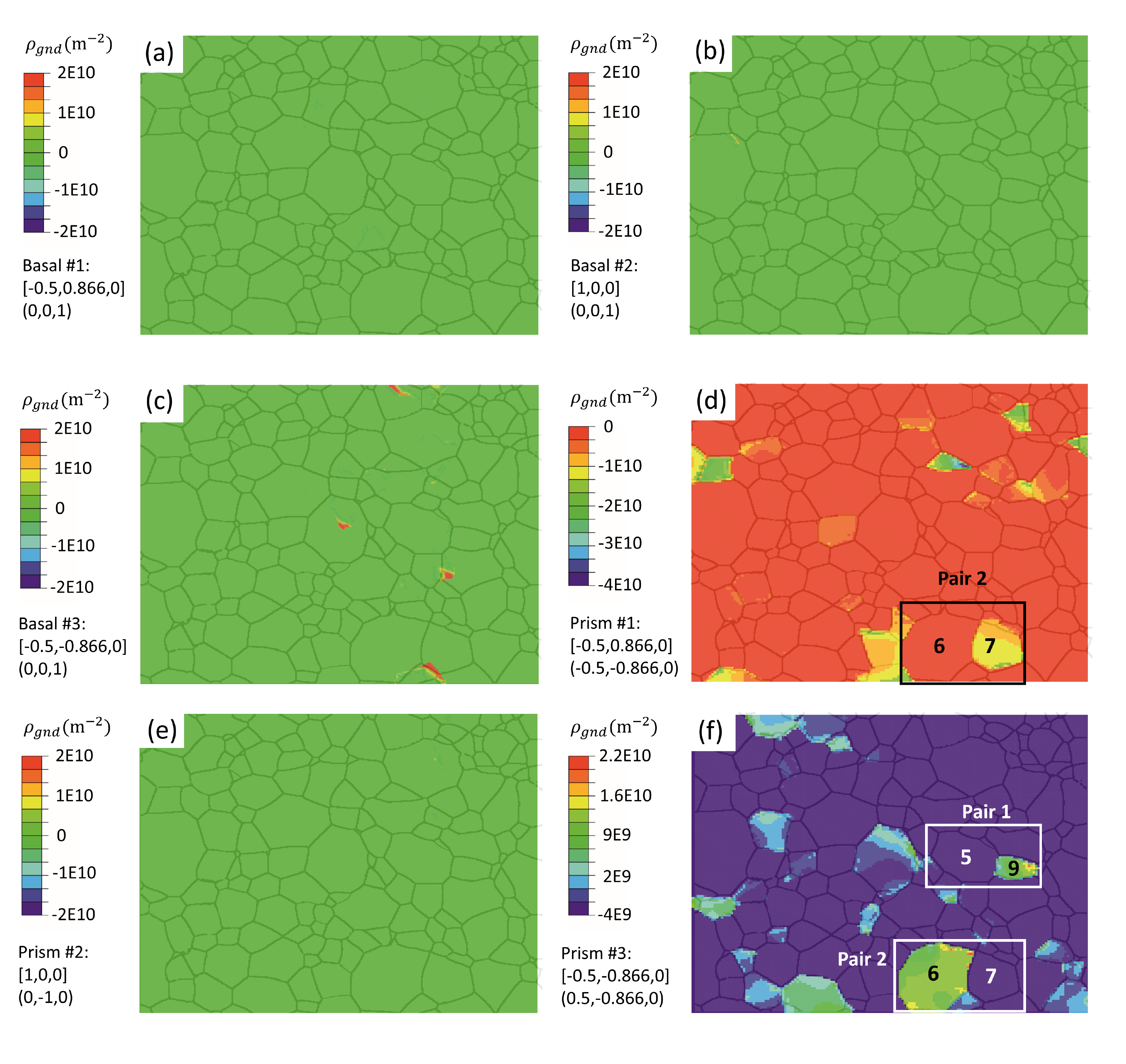}
\caption{Slip activity of prism and basal slip in sample 4 at room temperature, clear contrast were observed in (d) and (e), where Prism \#1 was activated in grain 7 and Prism \#3 slip was activated in grain 6 and grain 9.}
\label{fig.9}
\end{figure}

\clearpage

\begin{figure}[H]
\centering
\includegraphics[width=1\textwidth]{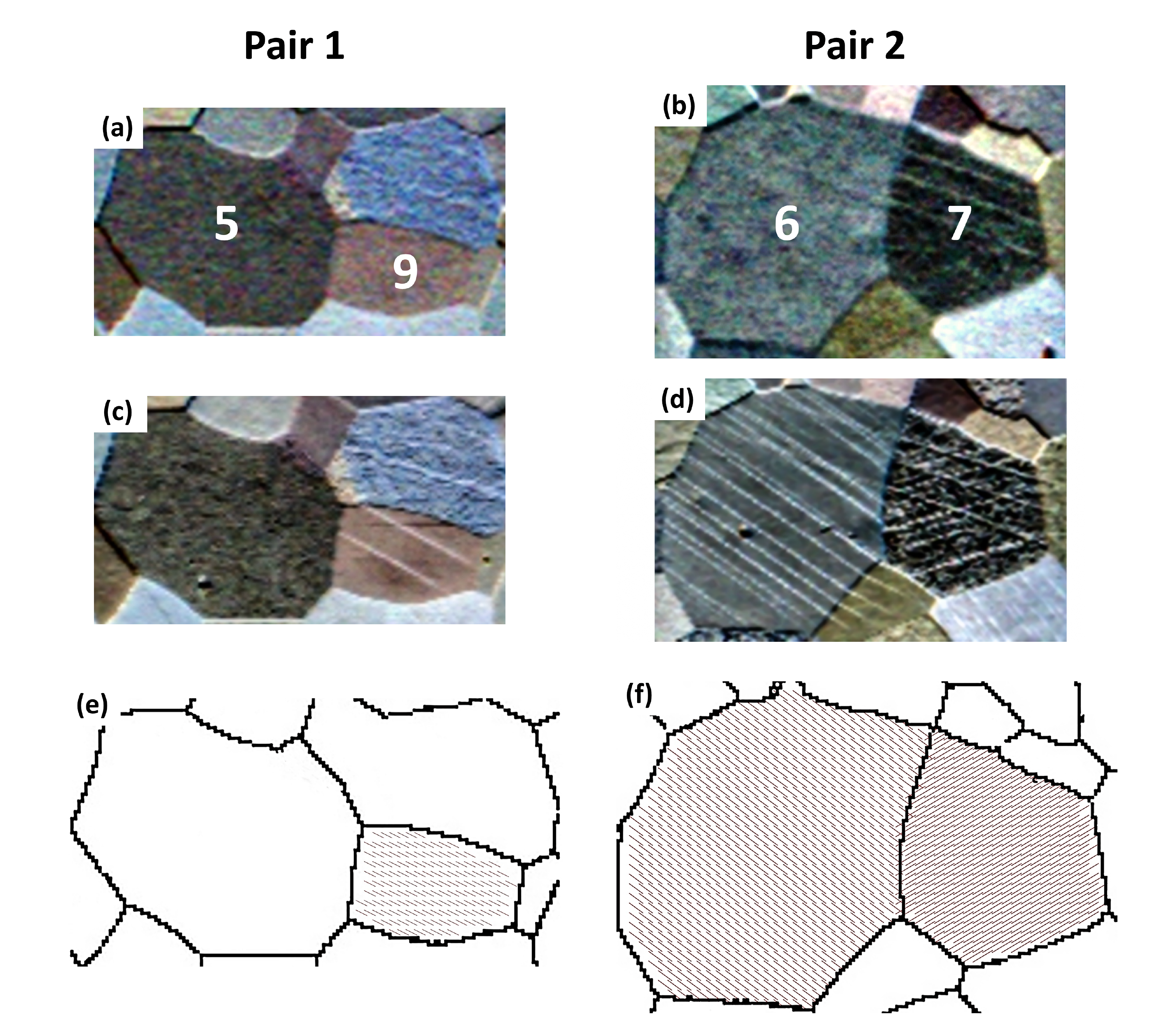}
\caption{Images of the two grain pairs before and after the creep test and simulated slip traces from the CPFE model.}
\label{fig.10}
\end{figure}

\clearpage

\begin{figure}[H]
\centering
\includegraphics[width=0.8\textwidth]{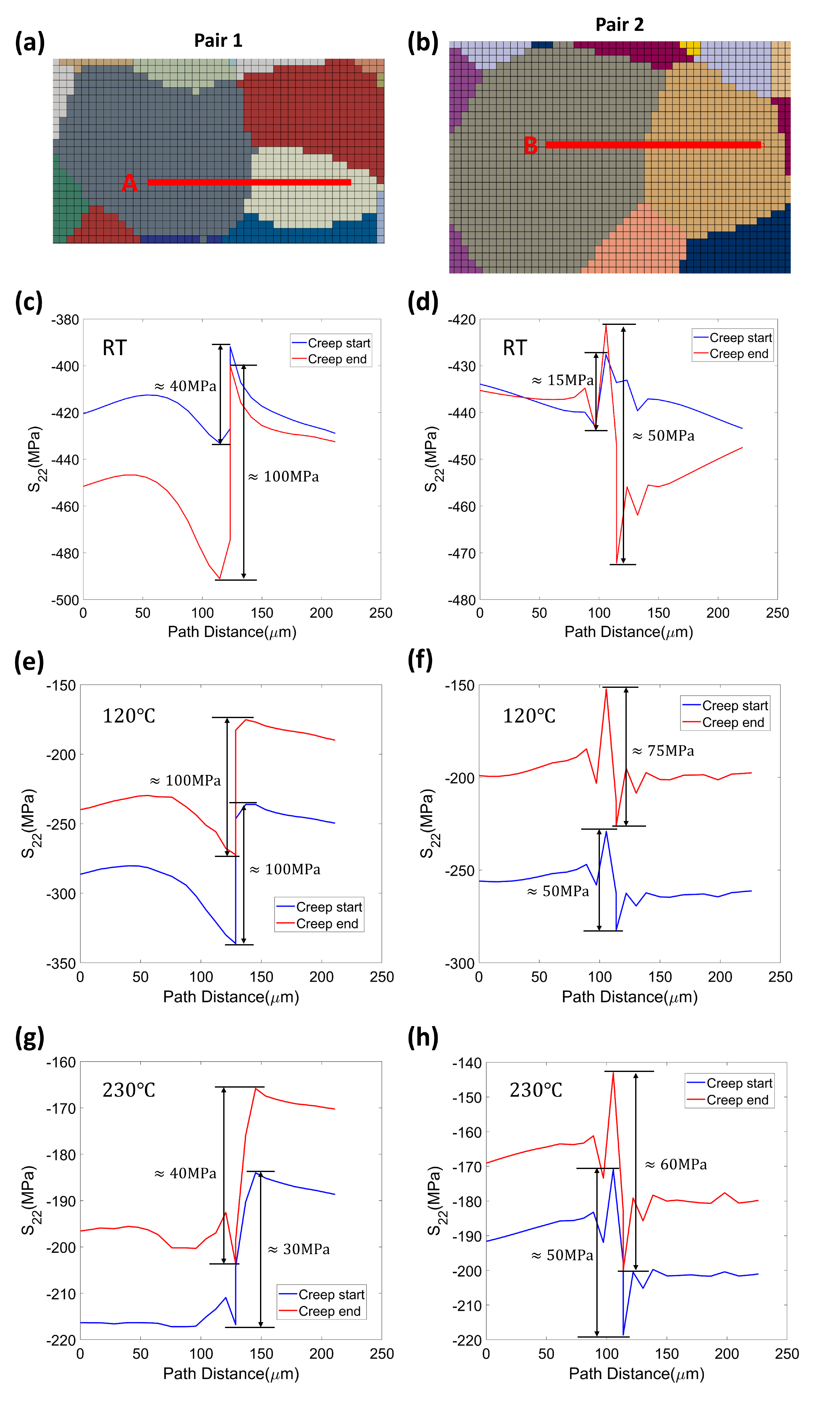}
\phantomcaption
\end{figure}
\begin{figure}[t!]\ContinuedFloat 
\centering
\includegraphics[width=0.75\textwidth]{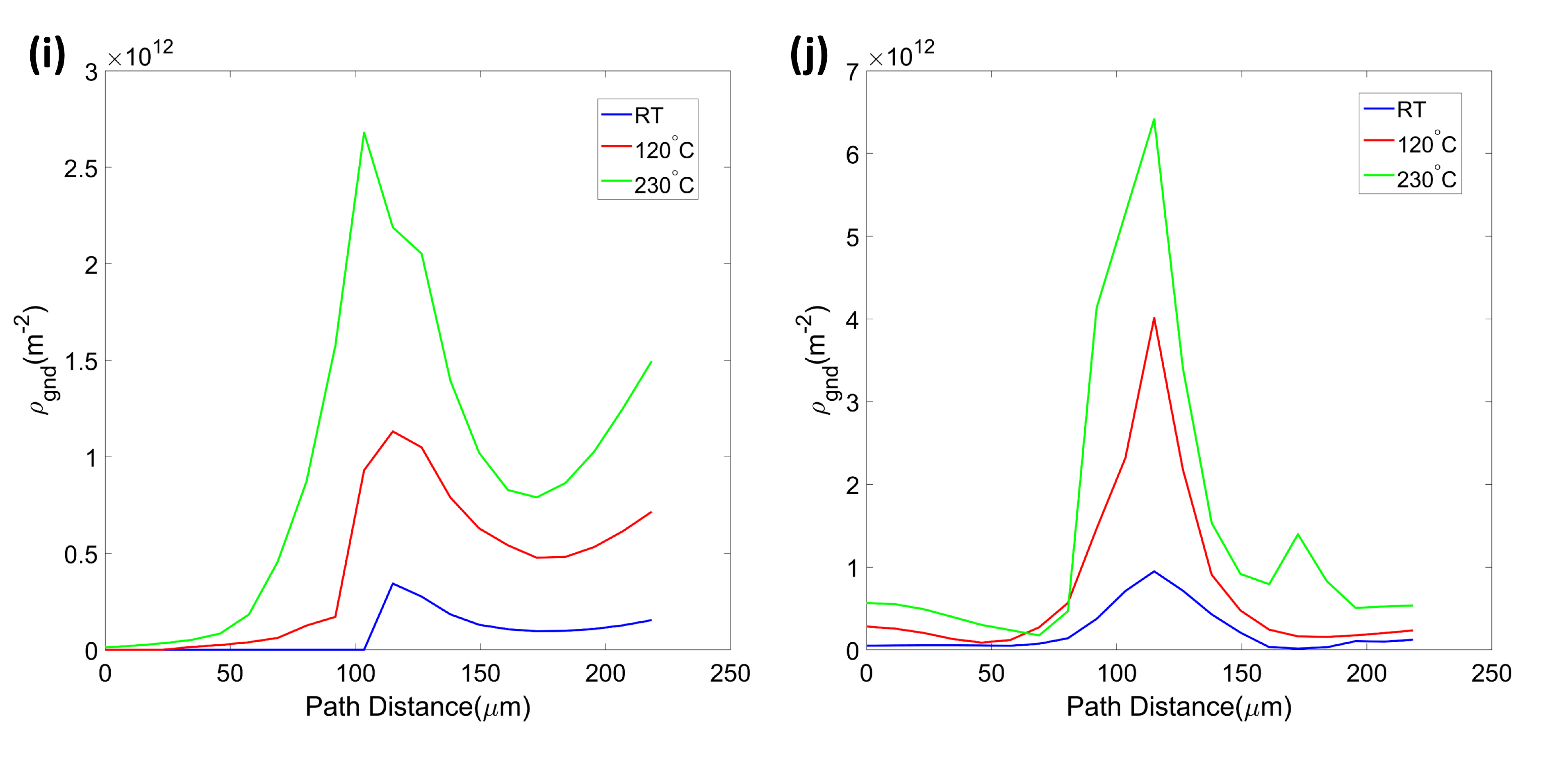}
\caption{Path within the grain pairs of (a) pair 1 (path A); (b) pair 2 (path B); S$_{22}$ component (along the loading direction) of (c) path A at room temperature; (d) path B at room temperature; (e) path A at 120~$^{\circ}$C; (f) path B at 120~$^{\circ}$C; (g) path A at 230~$^{\circ}$C; (h) path B at 230~$^{\circ}$C; GND density at the end of creep of (i) path A and (j) path B. (Note that the vertical axes are not the same in each plot).}
\label{fig.11}
\end{figure}

\section{Discussion}

The CRSS values obtained in this work are in sound agreement with the values reported in literature~\cite{HASIJA2003,ZHANG2015,Williams2002,BRITTON2015,SAKAI19741359,SAKAI1974545}, but are perhaps slightly lower particularly at elevated temperatures. Two reasons for these difference could be: (1) the Ti6Al alloy tested in this work had lower Al content ($\approx$ 5.8wt\%) compared to that in the literature, (2) strain rates in our work extend to lower values. Aluminium, as a common substitutional alloying addition to Ti, has proved to have strengthening effect on both basal and prism slip~\cite{Williams2002,SAKAI19741359,SAKAI1974545,Sakai1974FailureOS}. The strengthening is more marked for prism slip however so that the strengths of prism and basal slip systems are more similar in Ti6Al than in pure Ti. 

\begin{figure}[hbt!]
\centering
\includegraphics[width=0.8\textwidth]{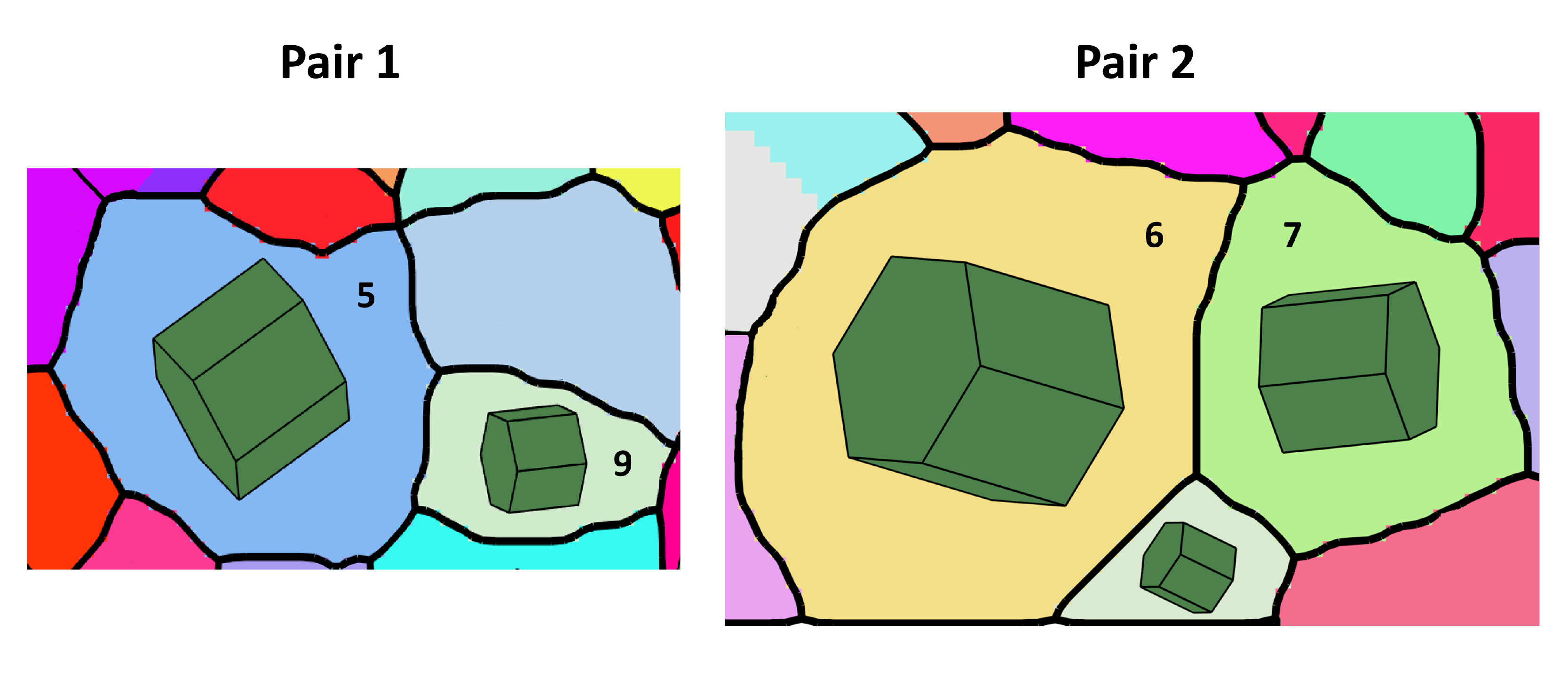}
\caption{Grain orientations of the two grain pairs.}
\label{fig.12}
\end{figure}

The grain orientations of the two grain pairs are shown in Figure~\ref{fig.12}. The angles between the $c$-axis and the loading direction were 56$^{\circ}$ for grain 5, 83$^{\circ}$ for grain 9, 74$^{\circ}$ for grain 6 and 73$^{\circ}$ for grain 7. Although the grain pair 1 behaved as a ‘rogue’ grain pair in this experiment, it is not a typical hard-soft grain combination. In grain pair 1, grain 9 was in ‘soft’ orientation but grain 5 was not a typical ‘hard’ grain and in fact it yields in the simulations for 120$^{\circ}$C and 230$^{\circ}$C. This is consistent with the observation by Brandes \textit{et al}~\cite{BRANDES2010}, where dwell fatigue facet was observed in a material lacking ‘hard’ orientated grains (with their $c$-axis nearly parallel to the loading direction). These non-typical ‘rogue’ grain pair could also lead to dwell fatigue failure. Due to the stress level being quite low to activate basal slip in this experiment, grains oriented favourably for prism slip systems were classified as ‘soft’ grains. Xu \textit{et al}~\cite{XU2020} suggested that the cold dwell fatigue and the dwell debit is sensitive to the dwell stress, which agrees well with the observations in this work. The origin of the dwell debit is the activation of slip systems, resulting in an accumulation of plastic strain during the stress dwell, which leads to load shedding from ‘soft’ grains to neighbouring ‘hard’ grains. High stress difference at the soft-hard grain boundary and stress increase in the ‘hard’ grain were observed and believed to lead to a nucleation of facets within the ‘hard’ grain~\cite{HASIJA2003,Sinha2006_1,BACHE2010}. As external stress changes, the activation of slip systems changes, thus the definition of ‘soft’ and ‘hard’ grains will vary dynamically. In reality the binary hard-soft nomenclature is in fact a continuum of varied relative grain strengths which are affected by external conditions such as temperature and macroscopic stress level. 

Schmid factor analysis of these grains can be found in Table.~\ref{table.5}. The Schmid factor of the active slip systems are marked as red in the table. It is found that the actual slip systems activated were not always the ones with the highest Schmid factor. This implies that the local stress states is different from the macroscopic stress state arising from the loading direction on the sample itself. The activation of slip systems is thus not always dependent on only that particular grain orientation with respect to the macroscopic load direction, but the local stress states plays a more crucial role. As a result, predicting the ‘hard’ and ‘soft’ grains based on macroscopic load direction and grain orientations in polycrystal samples is not always viable.

In Figure~\ref{fig.10}, grain 6 and the grain at its bottom right have very similar crystallographic orientations (refer to Figure~\ref{fig.1} along the loading direction). Slip transfers easily across the boundaries between these similar orientated grains, which behave as if the boundary between them was not present. Partially correlated group of grains often arrive in Ti alloys and are known as macrozones and increase the effective grain or structural unit size~\cite{Germain2008,BIAVANT2002}. Creep can happen in a ‘soft’ macrozone that is orientated favourably for slip, which leads to load shedding to a neighbouring ‘hard’ macrozone orientated poorly for slip and crack can initiate within the ‘hard’ macrozone~\cite{BANTOUNAS2010,ZHANGK2015,ZHANGK2017}. 

\begin{table}[H]
\centering
\caption{Schmid factor of the three prism slip systems in the two grain pairs, the active slip systems were marked as red.}
 \begin{tabular*}{0.75\textwidth}{ c@{\extracolsep{\fill}}c@{\extracolsep{\fill}}c@{\extracolsep{\fill}}c@{\extracolsep{\fill}}c} 
 \hline
 & Grain 5  & Grain 9 & Grain 6 & Grain 7 \\ \hline
 Prism 1 & 0.026 & 0.356 & 0.374 & \textcolor{red}{0.299} \\ 
 Prism 2 & 0.095 & 0.170 & 0.110 & 0.024 \\
 Prism 3 & 0.121 & \textcolor{red}{0.185} & \textcolor{red}{0.264} & 0.061 \\
 \hline
\end{tabular*}
\label{table.5}
\end{table}

\begin{figure}[hbt!]
\centering
\includegraphics[width=1\textwidth]{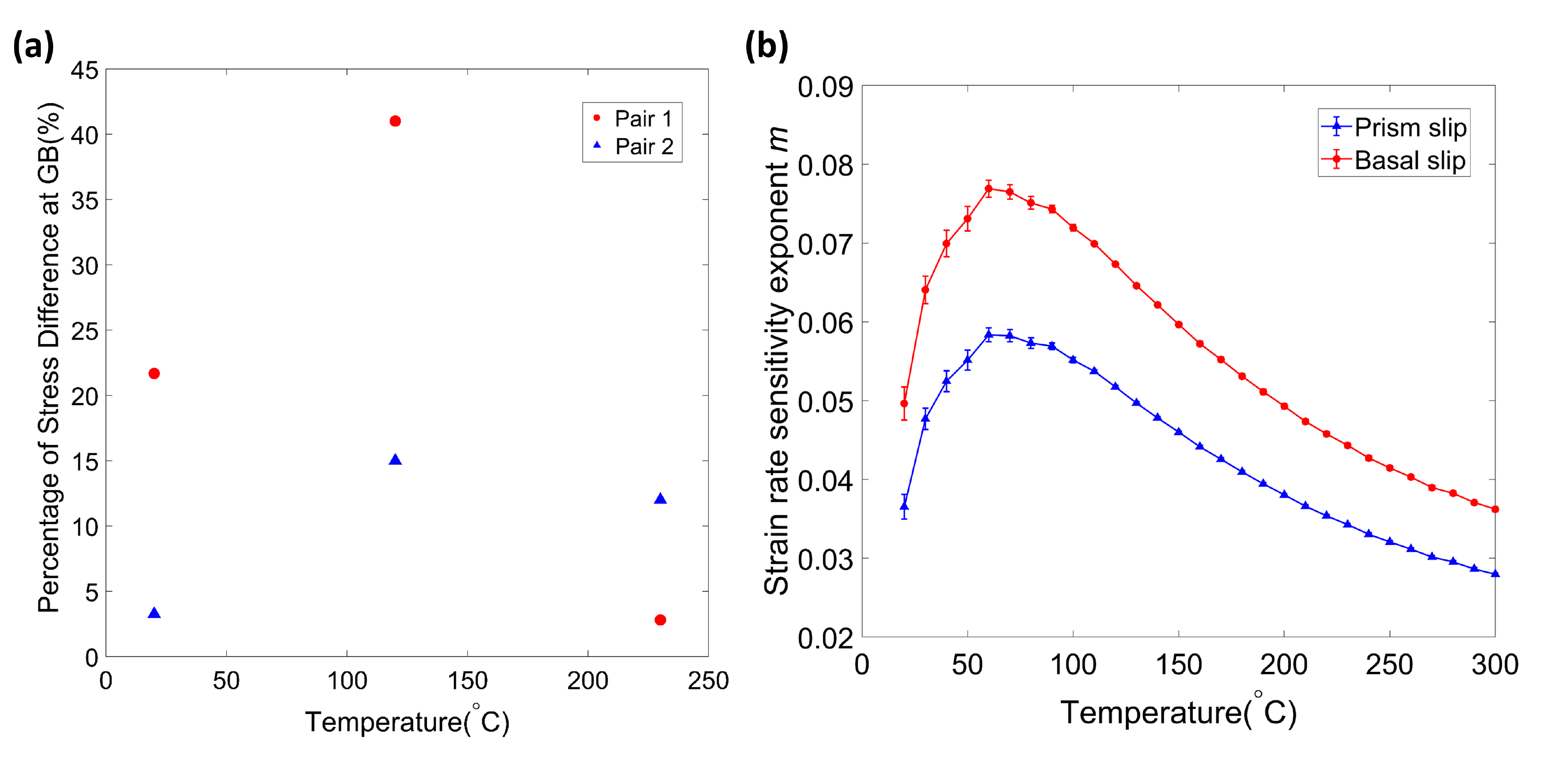}
\caption{(a) Percentage of stress difference at the grain boundaries of the two grain pairs at the end of the creep vs. temperature; (b) Strain rate sensitivity of basal and prism slip vs. temperature.}
\label{fig.13}
\end{figure}

Load shedding was found to be sensitive to temperature. As the energy required to activate basal and prismatic slip decreases when temperature increases~\cite{XIONG2020_2}, more plasticity was activated at higher temperatures. Stresses were found to relax in both ‘rogue’ and ‘non-rogue’ grain pairs. Although the stresses were relaxed, the differences in stress at the grain boundaries still exist. Since the stress level in each grain pair was different, direct comparison of the stress difference was not possible. Instead, the percentage of stress difference was calculated by dividing the averaged stress along the paths. Figure~\ref{fig.13}(a) shows the percentage of stress difference at the grain boundaries at the end of the creep against temperature. At room temperature and 120~$^{\circ}$C, stress differences in grain pair 1 were significantly higher than that in grain pair 2. At 230~$^{\circ}$C, the stress differences in the two pairs altered but both of them reduce to relatively low levels. It was found that in both grain pairs, the percentage stress difference increased as temperature increased to 120~$^{\circ}$C and decreased as the temperature further increased to 230~$^{\circ}$C. Among the three temperature assessed, 120~$^{\circ}$C was found to be the worst case scenario. To reveal the mechanism leading to this observation, the activity of slip systems need to be quantified. 

Strain rate sensitivity (SRS) are often used to express the slip activity in literature and it is considered to be a significant factor that controls load shedding~\cite{JUN2016NANO}. The strain rate sensitivity exponent, $m$, is calculated by: 
\begin{equation}
    m=d(log(\sigma))/d(log(\dot{\varepsilon}))
\end{equation}
which corresponds to the gradient of a $log(\sigma)$ vs. $log(\dot{\varepsilon})$ plot. The $m$ value is crystallographic orientation dependent~\cite{BRITTON2015}, where $m$ values of ‘soft’ grains are typically higher than that of ‘hard’ grains at room temperature~\cite{JUN2016NANO,XIONG2020}. This phenomenon is deemed to be related to the different SRS in different slip systems, which is confirmed by Jun \textit{et al}~\cite{JUN2016}, who reported that in $\alpha$-Ti prism slip had higher SRS ($m=0.07$) over basal slip ($m=0.03$) at room temperature. A more recent study~\cite{XIONG2020_2} shows that the SRS is also temperature dependent and higher SRS of a slip system leads to more creep strain accumulation during a stress dwell (i.e. higher activity of the slip system). We attempted to evaluate the SRS for the two major slip systems (prism and basal) at different temperatures in order to interpret the worst case scenario temperature. 

The three temperature dependent key parameters $\Delta{V}$, $\Delta{F}$ and $\tau_c$ were fitted with quadratic, linear and power law  functions~\cite{ZHANG2015,XIONG2020_2,ZHENG2018,TANAKA2016,KISHIDA2020168,MITCHELL20019827} so that the values can be predicted at different temperatures other than experimentally tested (recall Figure~\ref{fig.6} and Figure~\ref{fig.8}). The three parameters were calculated for every 10~$^{\circ}$C between the range of 20~$^{\circ}$C to 300~$^{\circ}$C and were inserted into the slip law (recall equation.~\ref{eqn:sliprateforwardbackward}). This allowed applied shear stress, $\tau_\alpha$ against shear strain rate, $\dot{\gamma}_\alpha$ to be plotted at log scale over a strain rate range of 10$^{-7}$ to 10$^{-5}$ s$^{-1}$ (the creep strain rate range observed in this work). Linear fitting was performed on the $\dot{\gamma}_\alpha$ vs. $\tau_\alpha$ curves and the gradients were obtained to represent the $m$ values. The linear fitting errors are shown by the error bars. The strain rate sensitivity variation against temperature is shown in Figure~\ref{fig.13}(b). More details can be found in the previous work~\cite{XIONG2020_2}. Prism slip has higher SRS over basal slip between room temperature and 300~$^{\circ}$C. The SRS as a function of temperature for both prism and basal slip follow the same trend as the stress difference at grain boundaries (see Figure~\ref{fig.13}(a)). The peak values of $m$ for both prism and basal slip can be expected at approximately 80~$^{\circ}$C. The peak temperature is in remarkable agreement with the worst case scenario temperature for dwell debit from the experience of aero engine industry (between 90~$^{\circ}$C and 120~$^{\circ}$C)~\cite{ZHANG2015,ZHENG2017Mechanistic}. The worst dwell debit was caused by the highest SRS of the two major slip systems (i.e. the highest slip activities). During stress dwell at the worst case scenario temperature, high slip activity results in a greater creep strain accumulation in the ‘soft’ grains which in turn cause dislocation pile-up at the soft-hard grain boundaries and load shedding which generates a high stress difference at the grain boundaries. At higher temperatures (e.g. above 200~$^{\circ}$C), the SRS of both prism and basal slip systems reduced to lower levels compared to room temperature. The slip activity of the two major slip systems were low at higher temperatures, which explained the phenomenon that cold dwell fatigue diminished at temperatures above 230~$^{\circ}$C~\cite{ZHANG2015}.     

\section{Summary}

This work utilised DIC technique and CPFE modelling to study the temperature effect on cold dwell fatigue in Ti6Al alloy. CRSS of the two major slip systems (prism and basal) of $\alpha$-Ti were quantified as a function of temperature by calibrating the CPFE model from experimentally measured strain evolution in individual grains during creep. Load shedding was clearly observed in the ‘rogue’ grain pair, resulting in an increase in stress within the ‘hard’ grain and high stress difference at the grain boundary. As temperature increased, the stress difference became larger at 120~$^{\circ}$C. At higher temperatures (above 230~$^{\circ}$C), the stress difference dropped to a reasonably low level.   

Grain orientation analyses showed that the ‘soft’ and the ‘hard’ grain would dynamically vary depending on the applied stress and operating temperature, which results from the activation of different slip systems. Local stress state plays a more important role in the activation of slip systems compared to grain orientations with respect to macroscopic load direction in large polycrystals.

The strain rate sensitivity for prism slip is higher than that of basal at temperatures between room temperature and 300~$^{\circ}$C. The SRS of both prism and basal slip show an increase-and-decrease trend as temperature increases. The peak SRS is expected to be at approximately 80~$^{\circ}$C for both prism slip and basal slip. The worst case scenario temperature for dwell debit (between 90~$^{\circ}$C and 120~$^{\circ}$C) can be explained by the highest SRS and slip activity of the two major slip systems at these temperatures.

\section*{Data Availability}
The EBSD maps and DIC images recorded during this experiment will be made openly available on the website https://zenodo.org/. Scripts for post-processing are openly available on Github: https://github.com/sann6001

\section*{Author Contributions}
\textbf{Yi Xiong}: Data Curation, Formal Analysis, Methodology, Investigation, Software, Validation, Visualisation, Writing – Original Draft, Writing – Review \& Editing

\textbf{Nicol\`{o} Grilli}: Data Curation, Formal Analysis, Methodology, Software, Writing – Original Draft, Writing – Review \& Editing

\textbf{Phani S Karamched}: Data Curation, Investigation, Methodology, Software, Validation, Writing – Review \& Editing

\textbf{Bo-Shiuan Li}: Investigation, Methodology, Writing – Review \& Editing

\textbf{Edmund Tarleton}: Data Curation, Formal Analysis, Software, Supervision, Writing – Review \& Editing

\textbf{Angus J Wilkinson}: Conceptualization, Funding Acquisition, Methodology, Investigation, Project Administration, Supervision, Visualisation, Writing – Review \& Editing

\section*{Acknowledgements}
 We are grateful for use of characterisation facilities within the David Cockayne Centre for Electron Microscopy, Department of Materials, University of Oxford, which has benefitted from financial support provided by the Henry Royce Institute (Grant ref EP/R010145/1). YX expresses gratitude to the financial support of China Scholarship Council (CSC) and ET acknowledges EPSRC for support through Fellowship grant (EP/N007239/1).
 
\section*{Appendix A: Justification of the depth of the model}

\renewcommand{\thefigure}{A\arabic{figure}}
\setcounter{figure}{0}

\begin{figure}[H]
\centering
\includegraphics[width=0.8\textwidth]{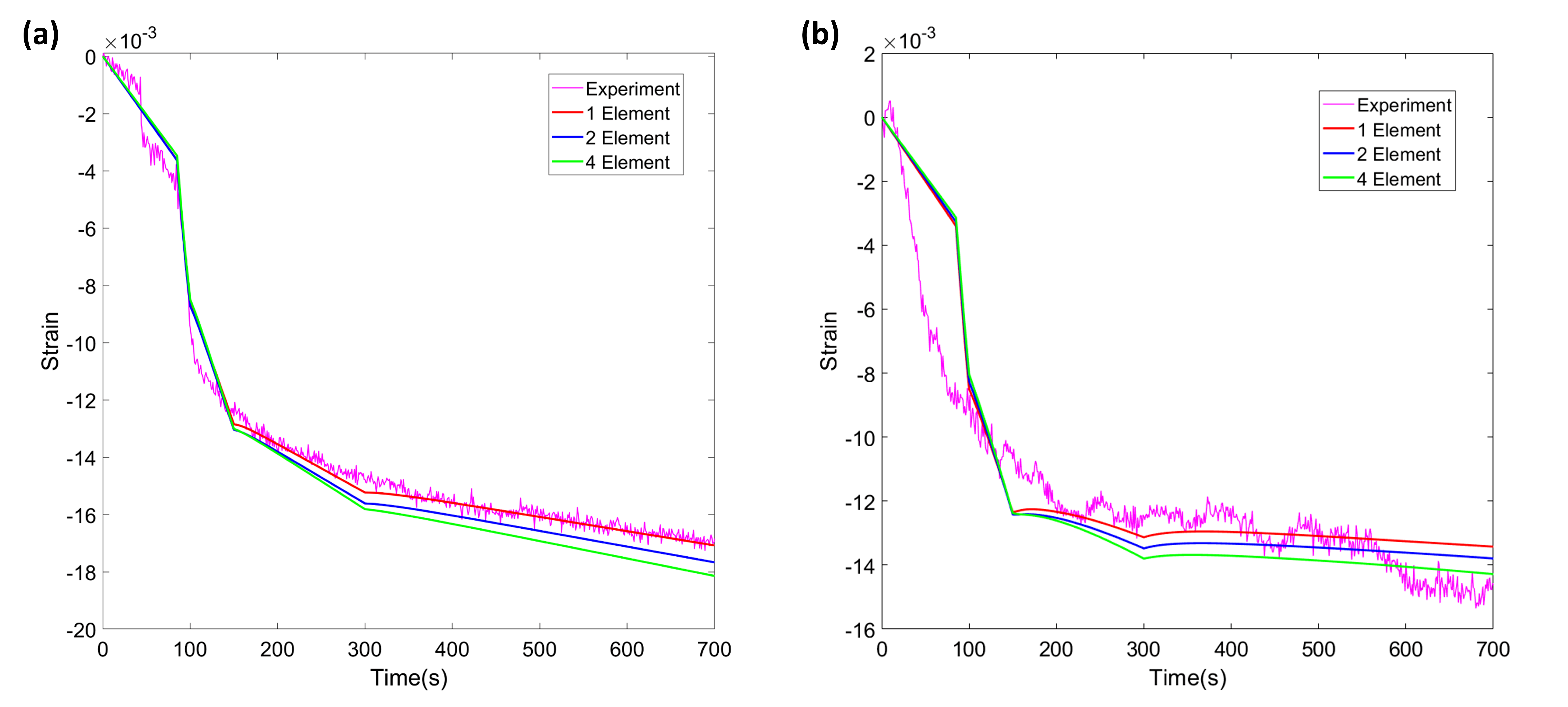}
\caption{Comparison of the strain in the models with 1 element, 2 elements and 4 elements in depth and the experimental strain of (a) grain 5 and (b) grain 7 in sample 1.}
\label{fig.a1}
\end{figure}

Three models with depth of 1 layer of elements, 2 layers of elements and 4 layers of elements were created for sample 1. Simulations were performed using same set of parameters (not the final optimised parameters). Figure~\ref{fig.a1} shows the comparison of the strains of the grain 5 and the grain 7 from the three models and experimental results. It is found that as the depth of the models increased from 1 element to 4 element, the simulated strain decreased. However, the difference in strains of the three models is relatively small, which did not affect their fittings with the experimental strains. To ensure enough degree of freedom of the elements on the top layer (where simulation results output from) and efficiency of the simulations, the model depth was set to be 2 elements for simulations presented in this paper.       

\section*{Appendix B: Macroscopic stress}

\renewcommand{\thefigure}{B\arabic{figure}}
\setcounter{figure}{0}

\begin{figure}[hbt!]
\centering
\includegraphics[width=0.8\textwidth]{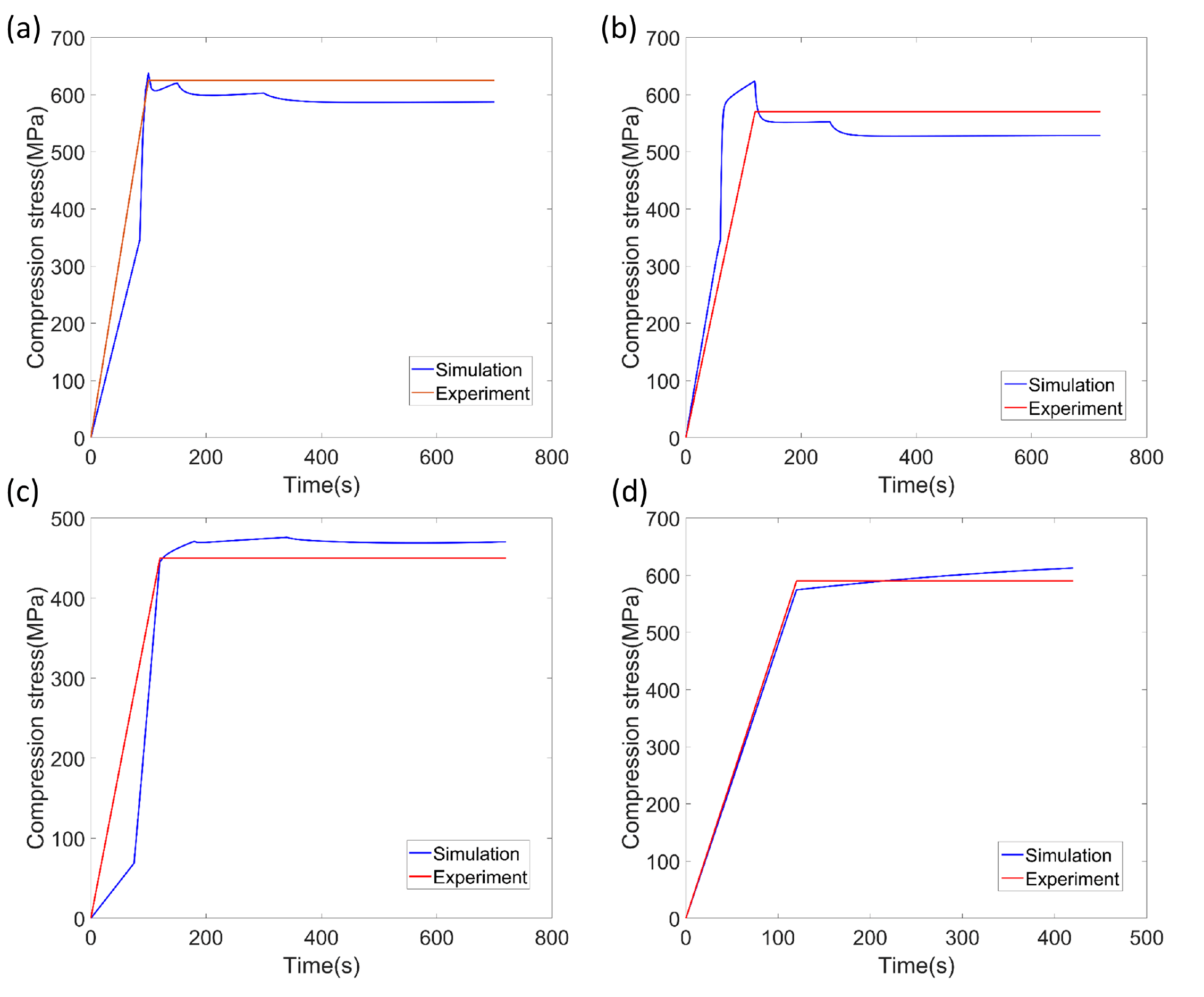}
\caption{Comparison of the macroscopic stresses of experiments and simulations for (a) sample 1 at room temperature; (b) sample 2 at 120$^{\circ}$C; (c) sample 3 at 230$^{\circ}$C and (d) sample 4 at room temperature.}
\label{fig.b1}
\end{figure}

Figure~\ref{fig.b1} shows the comparison of macroscopic stresses between experiments and simulations for the four samples. The experimental stress was estimated by dividing the initial sample cross-section area from the force recorded by the mechanical test frame. In simulations, forces were applied by displacement and we simplified the curved strain vs. time curves with several straight line segments (recall Figure~\ref{fig.4}(a)-(c)). This results in the observation of stress relaxation periods in the simulation stress, particularly for sample 1 and sample 2. Generally speaking, the macroscopic stress for experiments and simulations are in good agreement.  

\section*{Appendix C: Fitting errors and optimisation efficiency}

\renewcommand{\thetable}{C\arabic{table}}
\setcounter{table}{0}

The fitting errors and number of iterations are summarised in Table~\ref{tab:c1}. The optimisation procedures were performed on a high function PC with an Intel Xeon(R) CPU E5-2687W v2 @ 3.40GHz with 8 cores. It took approximately 80 minutes for one iteration of the Nelder-Mead method to complete. Therefore, the completion of the optimisation procedure for one sample could take 2-4 weeks. Using the automatic optimisation tool could save time for people but could also occupy the computer resources. The choice of the initial parameters that input into the Nelder-Mead method is extremely important. Sample 4, for example, using the calibrated CRSS values as input has the lowest averaged error among these four samples. The efficiency of optimisation could be improved by a more accurate estimation of the materials properties in our case. More advanced machine learning method~\cite{Ozaki2017} could also be applied to improve the efficiency. 

\begin{table*}[htb]
\centering
\caption{Averaged difference in strain between the simulations and experiments of the 8 selected grains in four samples and approximate number of iterations of the Nelder-Mead method.}
\def\arraystretch{1}
\begin{tabular}{|l|c|c|}
\hline
 & Averaged errors & No. of iterations \\
\hline
Sample 1 & 4.3\% & $\approx$300 \\
\hline
Sample 2 & 7.6\% & $\approx$500 \\
\hline
Sample 3 & 6.8\% & $\approx$500\\
\hline
Sample 4 & 3.1\% & NA\\
\hline
\end{tabular}
\label{tab:c1}
\end{table*}

\clearpage
\bibliographystyle{elsarticle-num-names}
\bibliography{ref.bib}

\begin{thebibliography}{57}
\expandafter\ifx\csname natexlab\endcsname\relax\def\natexlab#1{#1}\fi
\providecommand{\url}[1]{\texttt{#1}}
\providecommand{\href}[2]{#2}
\providecommand{\path}[1]{#1}
\providecommand{\DOIprefix}{doi:}
\providecommand{\ArXivprefix}{arXiv:}
\providecommand{\URLprefix}{URL: }
\providecommand{\Pubmedprefix}{pmid:}
\providecommand{\doi}[1]{\href{http://dx.doi.org/#1}{\path{#1}}}
\providecommand{\Pubmed}[1]{\href{pmid:#1}{\path{#1}}}
\providecommand{\bibinfo}[2]{#2}
\ifx\xfnm\relax \def\xfnm[#1]{\unskip,\space#1}\fi
\bibitem[{Anahid et~al.(2011)Anahid, Samal, and Ghosh}]{ANAHID2011}
\bibinfo{author}{M.~Anahid}, \bibinfo{author}{M.~K. Samal},
  \bibinfo{author}{S.~Ghosh},
\newblock \bibinfo{title}{Dwell fatigue crack nucleation model based on crystal
  plasticity finite element simulations of polycrystalline titanium alloys},
\newblock \bibinfo{journal}{Journal of the Mechanics and Physics of Solids}
  \bibinfo{volume}{59} (\bibinfo{year}{2011}) \bibinfo{pages}{2157 -- 2176}.
  \URLprefix
  \url{http://www.sciencedirect.com/science/article/pii/S0022509611001062}.
  \DOIprefix\doi{10.1016/j.jmps.2011.05.003}.
\bibitem[{Bache(2003)}]{BACHE2003}
\bibinfo{author}{M.~Bache},
\newblock \bibinfo{title}{A review of dwell sensitive fatigue in titanium
  alloys: the role of microstructure, texture and operating conditions},
\newblock \bibinfo{journal}{International Journal of Fatigue}
  \bibinfo{volume}{25} (\bibinfo{year}{2003}) \bibinfo{pages}{1079 -- 1087}.
  \URLprefix
  \url{http://www.sciencedirect.com/science/article/pii/S0142112303001452}.
  \DOIprefix\doi{10.1016/S0142-1123(03)00145-2}, \bibinfo{note}{international
  Conference on Fatigue Damage of Structural Materials IV}.
\bibitem[{Tympel et~al.(2016)Tympel, Lindley, Saunders, Dixon, and
  Dye}]{TYMPEL2016}
\bibinfo{author}{P.~Tympel}, \bibinfo{author}{T.~Lindley},
  \bibinfo{author}{E.~Saunders}, \bibinfo{author}{M.~Dixon},
  \bibinfo{author}{D.~Dye},
\newblock \bibinfo{title}{Influence of complex lcf and dwell load regimes on
  fatigue of {T}i–6{A}l–4{V}},
\newblock \bibinfo{journal}{Acta Materialia} \bibinfo{volume}{103}
  (\bibinfo{year}{2016}) \bibinfo{pages}{77 -- 88}. \URLprefix
  \url{http://www.sciencedirect.com/science/article/pii/S1359645415006801}.
  \DOIprefix\doi{10.1016/j.actamat.2015.09.014}.
\bibitem[{Ozturk et~al.(2017)Ozturk, Pilchak, and Ghosh}]{OZTURK2017}
\bibinfo{author}{D.~Ozturk}, \bibinfo{author}{A.~Pilchak},
  \bibinfo{author}{S.~Ghosh},
\newblock \bibinfo{title}{Experimentally validated dwell and cyclic fatigue
  crack nucleation model for $\alpha$–titanium alloys},
\newblock \bibinfo{journal}{Scripta Materialia} \bibinfo{volume}{127}
  (\bibinfo{year}{2017}) \bibinfo{pages}{15 -- 18}. \URLprefix
  \url{http://www.sciencedirect.com/science/article/pii/S1359646216304079}.
  \DOIprefix\doi{10.1016/j.scriptamat.2016.08.031}.
\bibitem[{Hasija et~al.(2003)Hasija, Ghosh, Mills, and Joseph}]{HASIJA2003}
\bibinfo{author}{V.~Hasija}, \bibinfo{author}{S.~Ghosh}, \bibinfo{author}{M.~J.
  Mills}, \bibinfo{author}{D.~S. Joseph},
\newblock \bibinfo{title}{Deformation and creep modeling in polycrystalline
  ti–6al alloys},
\newblock \bibinfo{journal}{Acta Materialia} \bibinfo{volume}{51}
  (\bibinfo{year}{2003}) \bibinfo{pages}{4533 -- 4549}. \URLprefix
  \url{http://www.sciencedirect.com/science/article/pii/S1359645403002891}.
  \DOIprefix\doi{10.1016/S1359-6454(03)00289-1}.
\bibitem[{Deka et~al.(2006)Deka, Joseph, Ghosh, and Mills}]{DEKA2006}
\bibinfo{author}{D.~Deka}, \bibinfo{author}{D.~S. Joseph},
  \bibinfo{author}{S.~Ghosh}, \bibinfo{author}{M.~J. Mills},
\newblock \bibinfo{title}{Crystal plasticity modeling of deformation and creep
  in polycrystalline ti-6242},
\newblock \bibinfo{journal}{Metallurgical and Materials Transactions A}
  \bibinfo{volume}{37} (\bibinfo{year}{2006}) \bibinfo{pages}{1371 -- 1388}.
  \URLprefix \url{https://doi.org/10.1007/s11661-006-0082-2}.
  \DOIprefix\doi{10.1007/s11661-006-0082-2}.
\bibitem[{Xu et~al.(2020)Xu, Joseph, Karamched, Fox, Rugg, Dunne, and
  Dye}]{XU2020}
\bibinfo{author}{Y.~Xu}, \bibinfo{author}{S.~Joseph},
  \bibinfo{author}{P.~Karamched}, \bibinfo{author}{K.~Fox},
  \bibinfo{author}{D.~Rugg}, \bibinfo{author}{F.~P.~E. Dunne},
  \bibinfo{author}{D.~Dye},
\newblock \bibinfo{title}{Predicting dwell fatigue life in titanium alloys
  using modelling and experiment},
\newblock \bibinfo{journal}{Nature Communications} \bibinfo{volume}{11}
  (\bibinfo{year}{2020}) \bibinfo{pages}{5868}. \URLprefix
  \url{https://doi.org/10.1038/s41467-020-19470-w}.
  \DOIprefix\doi{10.1038/s41467-020-19470-w}.
\bibitem[{Dunne et~al.(2007{\natexlab{a}})Dunne, Rugg, and
  Walker}]{DUNNE20071061}
\bibinfo{author}{F.~Dunne}, \bibinfo{author}{D.~Rugg},
  \bibinfo{author}{A.~Walker},
\newblock \bibinfo{title}{Lengthscale-dependent, elastically anisotropic,
  physically-based hcp crystal plasticity: Application to cold-dwell fatigue in
  ti alloys},
\newblock \bibinfo{journal}{International Journal of Plasticity}
  \bibinfo{volume}{23} (\bibinfo{year}{2007}{\natexlab{a}})
  \bibinfo{pages}{1061 -- 1083}. \URLprefix
  \url{http://www.sciencedirect.com/science/article/pii/S0749641906001641}.
  \DOIprefix\doi{10.1016/j.ijplas.2006.10.013}.
\bibitem[{Dunne et~al.(2007{\natexlab{b}})Dunne, Walker, and
  Rugg}]{DUNNE2007Proceedings}
\bibinfo{author}{F.~Dunne}, \bibinfo{author}{A.~Walker},
  \bibinfo{author}{D.~Rugg},
\newblock \bibinfo{title}{A systematic study of hcp crystal orientation and
  morphology effects in polycrystal deformation and fatigue},
\newblock \bibinfo{journal}{Proceedings of The Royal Society A: Mathematical,
  Physical and Engineering Sciences} \bibinfo{volume}{463}
  (\bibinfo{year}{2007}{\natexlab{b}}) \bibinfo{pages}{1467--1489}.
  \DOIprefix\doi{10.1098/rspa.2007.1833}.
\bibitem[{Sinha et~al.(2004)Sinha, Mills, and Williams}]{Sinha2004}
\bibinfo{author}{V.~Sinha}, \bibinfo{author}{M.~J. Mills},
  \bibinfo{author}{J.~C. Williams},
\newblock \bibinfo{title}{Understanding the contributions of normal-fatigue and
  static loading to the dwell fatigue in a near-alpha titanium alloy},
\newblock \bibinfo{journal}{Metallurgical and Materials Transactions A}
  \bibinfo{volume}{35} (\bibinfo{year}{2004}) \bibinfo{pages}{3141--3148}.
  \URLprefix \url{https://doi.org/10.1007/s11661-004-0058-z}.
  \DOIprefix\doi{10.1007/s11661-004-0058-z}.
\bibitem[{Sinha et~al.(2006{\natexlab{a}})Sinha, Mills, and
  Williams}]{Sinha2006_1}
\bibinfo{author}{V.~Sinha}, \bibinfo{author}{M.~J. Mills},
  \bibinfo{author}{J.~C. Williams},
\newblock \bibinfo{title}{Crystallography of fracture facets in a near-alpha
  titanium alloy},
\newblock \bibinfo{journal}{Metallurgical and Materials Transactions A}
  \bibinfo{volume}{37} (\bibinfo{year}{2006}{\natexlab{a}})
  \bibinfo{pages}{2015--2026}. \URLprefix
  \url{https://doi.org/10.1007/s11661-006-0144-5}.
  \DOIprefix\doi{10.1007/s11661-006-0144-5}.
\bibitem[{Sinha et~al.(2006{\natexlab{b}})Sinha, Mills, Williams, and
  Spowart}]{Sinha2006_2}
\bibinfo{author}{V.~Sinha}, \bibinfo{author}{M.~J. Mills},
  \bibinfo{author}{J.~C. Williams}, \bibinfo{author}{J.~E. Spowart},
\newblock \bibinfo{title}{Observations on the faceted initiation site in the
  dwell-fatigue tested ti-6242 alloy: Crystallographic orientation and size
  effects},
\newblock \bibinfo{journal}{Metallurgical and Materials Transactions A}
  \bibinfo{volume}{37} (\bibinfo{year}{2006}{\natexlab{b}})
  \bibinfo{pages}{1507--1518}. \URLprefix
  \url{https://doi.org/10.1007/s11661-006-0095-x}.
  \DOIprefix\doi{10.1007/s11661-006-0095-x}.
\bibitem[{Evans and Bache(1994)}]{EVANS1994}
\bibinfo{author}{W.~Evans}, \bibinfo{author}{M.~Bache},
\newblock \bibinfo{title}{Dwell-sensitive fatigue under biaxial loads in the
  near-alpha titanium alloy imi685},
\newblock \bibinfo{journal}{International Journal of Fatigue}
  \bibinfo{volume}{16} (\bibinfo{year}{1994}) \bibinfo{pages}{443 -- 452}.
  \URLprefix
  \url{http://www.sciencedirect.com/science/article/pii/0142112394901945}.
  \DOIprefix\doi{https://doi.org/10.1016/0142-1123(94)90194-5}.
\bibitem[{Brandes et~al.(2010)Brandes, Mills, and Williams}]{BRANDES2010}
\bibinfo{author}{M.~C. Brandes}, \bibinfo{author}{M.~J. Mills},
  \bibinfo{author}{J.~C. Williams},
\newblock \bibinfo{title}{The influence of slip character on the creep and
  fatigue fracture of an $\alpha$ ti-al alloy},
\newblock \bibinfo{journal}{Metallurgical and Materials Transactions A}
  \bibinfo{volume}{41} (\bibinfo{year}{2010}) \bibinfo{pages}{3463--3472}.
  \URLprefix \url{https://doi.org/10.1007/s11661-010-0407-z}.
  \DOIprefix\doi{10.1007/s11661-010-0407-z}.
\bibitem[{L{\"u}tjering and Williams(2003)}]{TITANIUM}
\bibinfo{author}{G.~L{\"u}tjering}, \bibinfo{author}{J.~Williams},
  \bibinfo{title}{Titanium}, Engineering materials and processes,
  \bibinfo{publisher}{Springer}, \bibinfo{year}{2003}. \URLprefix
  \url{https://books.google.co.uk/books?id=GwI9ul\_wAegC}.
\bibitem[{Zhang et~al.(2015)Zhang, Cuddihy, and Dunne}]{ZHANG2015}
\bibinfo{author}{Z.~Zhang}, \bibinfo{author}{M.~Cuddihy},
  \bibinfo{author}{F.~Dunne},
\newblock \bibinfo{title}{On rate-dependent polycrystal deformation: the
  temperature sensitivity of cold dwell fatigue},
\newblock \bibinfo{journal}{Proceedings of the Royal Society A: Mathematical,
  Physical and Engineering Science} \bibinfo{volume}{471}
  (\bibinfo{year}{2015}) \bibinfo{pages}{20150214}.
  \DOIprefix\doi{10.1098/rspa.2015.0214}.
\bibitem[{Bourigault et~al.(2016)Bourigault, Senn, and Eberl}]{PYDIC}
\bibinfo{author}{C.~Bourigault}, \bibinfo{author}{M.~Senn},
  \bibinfo{author}{C.~Eberl}, \bibinfo{title}{{Python DIC Software}},
  \bibinfo{year}{2016}. \URLprefix
  \url{https://github.com/phanikaramched/Python_DIC}.
\bibitem[{Kalidindi(1998)}]{KALIDINDI1998267}
\bibinfo{author}{S.~R. Kalidindi},
\newblock \bibinfo{title}{Incorporation of deformation twinning in crystal
  plasticity models},
\newblock \bibinfo{journal}{Journal of the Mechanics and Physics of Solids}
  \bibinfo{volume}{46} (\bibinfo{year}{1998}) \bibinfo{pages}{267 -- 290}.
  \URLprefix
  \url{http://www.sciencedirect.com/science/article/pii/S0022509697000513}.
  \DOIprefix\doi{https://doi.org/10.1016/S0022-5096(97)00051-3}.
\bibitem[{Grilli et~al.(2020)Grilli, Tarleton, Edmondson, Gussev, and
  Cocks}]{Grilli2020PRM}
\bibinfo{author}{N.~Grilli}, \bibinfo{author}{E.~Tarleton},
  \bibinfo{author}{P.~D. Edmondson}, \bibinfo{author}{M.~N. Gussev},
  \bibinfo{author}{A.~C.~F. Cocks},
\newblock \bibinfo{title}{In situ measurement and modelling of the growth and
  length scale of twins in $\ensuremath{\alpha}$-uranium},
\newblock \bibinfo{journal}{Phys. Rev. Materials} \bibinfo{volume}{4}
  (\bibinfo{year}{2020}) \bibinfo{pages}{043605}. \URLprefix
  \url{https://link.aps.org/doi/10.1103/PhysRevMaterials.4.043605}.
  \DOIprefix\doi{10.1103/PhysRevMaterials.4.043605}.
\bibitem[{Orowan(1934)}]{Orowan1934}
\bibinfo{author}{E.~Orowan},
\newblock \bibinfo{title}{Zur kristallplastizität. i},
\newblock \bibinfo{journal}{Zeitschrift für Physik} \bibinfo{volume}{89}
  (\bibinfo{year}{1934}) \bibinfo{pages}{605 -- 613}.
  \DOIprefix\doi{https://doi.org/10.1007/BF01341478}.
\bibitem[{Das et~al.(2018)Das, Hofmann, and Tarleton}]{DAS201818}
\bibinfo{author}{S.~Das}, \bibinfo{author}{F.~Hofmann},
  \bibinfo{author}{E.~Tarleton},
\newblock \bibinfo{title}{Consistent determination of geometrically necessary
  dislocation density from simulations and experiments},
\newblock \bibinfo{journal}{International Journal of Plasticity}
  \bibinfo{volume}{109} (\bibinfo{year}{2018}) \bibinfo{pages}{18 -- 42}.
  \URLprefix
  \url{http://www.sciencedirect.com/science/article/pii/S0749641918300068}.
  \DOIprefix\doi{https://doi.org/10.1016/j.ijplas.2018.05.001}.
\bibitem[{Roters et~al.(2019)Roters, Diehl, Shanthraj, Eisenlohr, Reuber, Wong,
  Maiti, Ebrahimi, Hochrainer, Fabritius, Nikolov, Friák, Fujita, Grilli,
  Janssens, Jia, Kok, Ma, Meier, Werner, Stricker, Weygand, and
  Raabe}]{ROTERS2019420}
\bibinfo{author}{F.~Roters}, \bibinfo{author}{M.~Diehl},
  \bibinfo{author}{P.~Shanthraj}, \bibinfo{author}{P.~Eisenlohr},
  \bibinfo{author}{C.~Reuber}, \bibinfo{author}{S.~Wong},
  \bibinfo{author}{T.~Maiti}, \bibinfo{author}{A.~Ebrahimi},
  \bibinfo{author}{T.~Hochrainer}, \bibinfo{author}{H.-O. Fabritius},
  \bibinfo{author}{S.~Nikolov}, \bibinfo{author}{M.~Friák},
  \bibinfo{author}{N.~Fujita}, \bibinfo{author}{N.~Grilli},
  \bibinfo{author}{K.~Janssens}, \bibinfo{author}{N.~Jia},
  \bibinfo{author}{P.~Kok}, \bibinfo{author}{D.~Ma},
  \bibinfo{author}{F.~Meier}, \bibinfo{author}{E.~Werner},
  \bibinfo{author}{M.~Stricker}, \bibinfo{author}{D.~Weygand},
  \bibinfo{author}{D.~Raabe},
\newblock \bibinfo{title}{Damask – the {D}üsseldorf advanced material
  simulation kit for modeling multi-physics crystal plasticity, thermal, and
  damage phenomena from the single crystal up to the component scale},
\newblock \bibinfo{journal}{Computational Materials Science}
  \bibinfo{volume}{158} (\bibinfo{year}{2019}) \bibinfo{pages}{420 -- 478}.
  \URLprefix
  \url{http://www.sciencedirect.com/science/article/pii/S0927025618302714}.
  \DOIprefix\doi{https://doi.org/10.1016/j.commatsci.2018.04.030}.
\bibitem[{Hill and Rice(1972)}]{HILL1972401}
\bibinfo{author}{R.~Hill}, \bibinfo{author}{J.~Rice},
\newblock \bibinfo{title}{Constitutive analysis of elastic-plastic crystals at
  arbitrary strain},
\newblock \bibinfo{journal}{Journal of the Mechanics and Physics of Solids}
  \bibinfo{volume}{20} (\bibinfo{year}{1972}) \bibinfo{pages}{401 -- 413}.
  \URLprefix
  \url{http://www.sciencedirect.com/science/article/pii/0022509672900178}.
  \DOIprefix\doi{https://doi.org/10.1016/0022-5096(72)90017-8}.
\bibitem[{Grilli et~al.(2020)Grilli, Cocks, and Tarleton}]{GRILLI2020109276}
\bibinfo{author}{N.~Grilli}, \bibinfo{author}{A.~C. Cocks},
  \bibinfo{author}{E.~Tarleton},
\newblock \bibinfo{title}{Crystal plasticity finite element modelling of
  coarse-grained $\alpha$-uranium},
\newblock \bibinfo{journal}{Computational Materials Science}
  \bibinfo{volume}{171} (\bibinfo{year}{2020}) \bibinfo{pages}{109276}.
  \URLprefix
  \url{http://www.sciencedirect.com/science/article/pii/S0927025619305750}.
  \DOIprefix\doi{https://doi.org/10.1016/j.commatsci.2019.109276}.
\bibitem[{Belytschko et~al.(2014)Belytschko, Liu, Moran, and
  Elkhodary}]{Belytschko2014}
\bibinfo{author}{T.~Belytschko}, \bibinfo{author}{K.~Liu},
  \bibinfo{author}{B.~Moran}, \bibinfo{author}{K.~Elkhodary},
  \bibinfo{title}{Nonlinear Finite Element Analysis for Continua and
  Structures}, \bibinfo{edition}{2} ed., \bibinfo{publisher}{John Wiley \&
  Sons}, \bibinfo{address}{New York}, \bibinfo{year}{2014}.
\bibitem[{Zhang et~al.(2016)Zhang, Jun, Britton, and Dunne}]{ZHANG2016393}
\bibinfo{author}{Z.~Zhang}, \bibinfo{author}{T.-S. Jun}, \bibinfo{author}{T.~B.
  Britton}, \bibinfo{author}{F.~P. Dunne},
\newblock \bibinfo{title}{Determination of ti-6242 $\alpha$ and $\beta$ slip
  properties using micro-pillar test and computational crystal plasticity},
\newblock \bibinfo{journal}{Journal of the Mechanics and Physics of Solids}
  \bibinfo{volume}{95} (\bibinfo{year}{2016}) \bibinfo{pages}{393 -- 410}.
  \URLprefix
  \url{http://www.sciencedirect.com/science/article/pii/S0022509616302307}.
  \DOIprefix\doi{https://doi.org/10.1016/j.jmps.2016.06.007}.
\bibitem[{Zheng et~al.(2016)Zheng, Balint, and Dunne}]{ZHENG2016411}
\bibinfo{author}{Z.~Zheng}, \bibinfo{author}{D.~S. Balint},
  \bibinfo{author}{F.~P. Dunne},
\newblock \bibinfo{title}{Dwell fatigue in two ti alloys: An integrated crystal
  plasticity and discrete dislocation study},
\newblock \bibinfo{journal}{Journal of the Mechanics and Physics of Solids}
  \bibinfo{volume}{96} (\bibinfo{year}{2016}) \bibinfo{pages}{411 -- 427}.
  \URLprefix
  \url{http://www.sciencedirect.com/science/article/pii/S002250961630223X}.
  \DOIprefix\doi{https://doi.org/10.1016/j.jmps.2016.08.008}.
\bibitem[{Xiong et~al.(2020)Xiong, Karamched, Nguyen, Collins, Grilli,
  Magazzeni, Tarleton, and Wilkinson}]{XIONG2020_2}
\bibinfo{author}{Y.~Xiong}, \bibinfo{author}{P.~S. Karamched},
  \bibinfo{author}{C.-T. Nguyen}, \bibinfo{author}{D.~M. Collins},
  \bibinfo{author}{N.~Grilli}, \bibinfo{author}{C.~M. Magazzeni},
  \bibinfo{author}{E.~Tarleton}, \bibinfo{author}{A.~J. Wilkinson},
  \bibinfo{title}{An in-situ synchrotron diffraction study of stress relaxation
  in titanium: Effect of temperature and oxygen on cold dwell fatigue},
  \bibinfo{year}{2020}. \href{http://arxiv.org/abs/2011.10041}{{\tt
  arXiv:2011.10041}}.
\bibitem[{Williams et~al.(2002)Williams, Baggerly, and Paton}]{Williams2002}
\bibinfo{author}{J.~C. Williams}, \bibinfo{author}{R.~G. Baggerly},
  \bibinfo{author}{N.~E. Paton},
\newblock \bibinfo{title}{Deformation behavior of hcp ti-al alloy single
  crystals},
\newblock \bibinfo{journal}{Metallurgical and Materials Transactions A}
  \bibinfo{volume}{33} (\bibinfo{year}{2002}) \bibinfo{pages}{837--850}.
  \URLprefix \url{https://doi.org/10.1007/s11661-002-0153-y}.
  \DOIprefix\doi{10.1007/s11661-002-0153-y}.
\bibitem[{Arsenlis and Parks(1999)}]{ARSENLIS19991597}
\bibinfo{author}{A.~Arsenlis}, \bibinfo{author}{D.~Parks},
\newblock \bibinfo{title}{Crystallographic aspects of geometrically-necessary
  and statistically-stored dislocation density},
\newblock \bibinfo{journal}{Acta Materialia} \bibinfo{volume}{47}
  (\bibinfo{year}{1999}) \bibinfo{pages}{1597 -- 1611}. \URLprefix
  \url{http://www.sciencedirect.com/science/article/pii/S1359645499000208}.
  \DOIprefix\doi{https://doi.org/10.1016/S1359-6454(99)00020-8}.
\bibitem[{Nye(1953)}]{NYE1953153}
\bibinfo{author}{J.~Nye},
\newblock \bibinfo{title}{Some geometrical relations in dislocated crystals},
\newblock \bibinfo{journal}{Acta Metallurgica} \bibinfo{volume}{1}
  (\bibinfo{year}{1953}) \bibinfo{pages}{153 -- 162}. \URLprefix
  \url{http://www.sciencedirect.com/science/article/pii/0001616053900546}.
  \DOIprefix\doi{https://doi.org/10.1016/0001-6160(53)90054-6}.
\bibitem[{Tarleton(2020)}]{CrystalPlasticityUMAT}
\bibinfo{author}{E.~Tarleton}, \bibinfo{title}{Crystalplasticity},
  \bibinfo{howpublished}{\url{https://github.com/TarletonGroup/CrystalPlasticity}},
  \bibinfo{year}{2020}.
\bibitem[{Grilli(2020)}]{EBSD2ABAQUS}
\bibinfo{author}{N.~Grilli}, \bibinfo{title}{{ebsd2abaqusEuler}},
  \bibinfo{year}{2020}. \URLprefix
  \url{https://github.com/ngrilli/ebsd2abaqusEuler}.
\bibitem[{Jones et~al.(2001)Jones, Oliphant, Peterson et~al.}]{SciPy}
\bibinfo{author}{E.~Jones}, \bibinfo{author}{T.~Oliphant},
  \bibinfo{author}{P.~Peterson}, et~al., \bibinfo{title}{{SciPy}: Open source
  scientific tools for {Python}}, \bibinfo{year}{2001}.
\bibitem[{Nelder and Mead(1965)}]{NelderMead1965}
\bibinfo{author}{J.~A. Nelder}, \bibinfo{author}{R.~Mead},
\newblock \bibinfo{title}{{A Simplex Method for Function Minimization}},
\newblock \bibinfo{journal}{The Computer Journal} \bibinfo{volume}{7}
  (\bibinfo{year}{1965}) \bibinfo{pages}{308--313}.
  \DOIprefix\doi{10.1093/comjnl/7.4.308}.
\bibitem[{Grilli et~al.(2020)Grilli, Earp, Cocks, Marrow, and
  Tarleton}]{GRILLI2020103800}
\bibinfo{author}{N.~Grilli}, \bibinfo{author}{P.~Earp}, \bibinfo{author}{A.~C.
  Cocks}, \bibinfo{author}{J.~Marrow}, \bibinfo{author}{E.~Tarleton},
\newblock \bibinfo{title}{Characterisation of slip and twin activity using
  digital image correlation and crystal plasticity finite element simulation:
  Application to orthorhombic $\alpha$-uranium},
\newblock \bibinfo{journal}{Journal of the Mechanics and Physics of Solids}
  \bibinfo{volume}{135} (\bibinfo{year}{2020}) \bibinfo{pages}{103800}.
  \URLprefix
  \url{http://www.sciencedirect.com/science/article/pii/S0022509619306696}.
  \DOIprefix\doi{https://doi.org/10.1016/j.jmps.2019.103800}.
\bibitem[{Kishida et~al.(2020)Kishida, Kim, Nagae, and Inui}]{KISHIDA2020168}
\bibinfo{author}{K.~Kishida}, \bibinfo{author}{J.~G. Kim},
  \bibinfo{author}{T.~Nagae}, \bibinfo{author}{H.~Inui},
\newblock \bibinfo{title}{Experimental evaluation of critical resolved shear
  stress for the first-order pyramidal $c+a$ slip in commercially pure ti by
  micropillar compression method},
\newblock \bibinfo{journal}{Acta Materialia} \bibinfo{volume}{196}
  (\bibinfo{year}{2020}) \bibinfo{pages}{168--174}. \URLprefix
  \url{https://www.sciencedirect.com/science/article/pii/S135964542030478X}.
  \DOIprefix\doi{https://doi.org/10.1016/j.actamat.2020.06.043}.
\bibitem[{Mitchell(2001)}]{MITCHELL20019827}
\bibinfo{author}{T.~Mitchell},
\newblock \bibinfo{title}{Yielding in crystals containing atomic-size
  obstacles},
\newblock in: \bibinfo{editor}{K.~J. Buschow}, \bibinfo{editor}{R.~W. Cahn},
  \bibinfo{editor}{M.~C. Flemings}, \bibinfo{editor}{B.~Ilschner},
  \bibinfo{editor}{E.~J. Kramer}, \bibinfo{editor}{S.~Mahajan},
  \bibinfo{editor}{P.~Veyssière} (Eds.), \bibinfo{booktitle}{Encyclopedia of
  Materials: Science and Technology}, \bibinfo{publisher}{Elsevier},
  \bibinfo{address}{Oxford}, \bibinfo{year}{2001}, pp.
  \bibinfo{pages}{9827--9842}. \URLprefix
  \url{https://www.sciencedirect.com/science/article/pii/B0080431526017812}.
  \DOIprefix\doi{https://doi.org/10.1016/B0-08-043152-6/01781-2}.
\bibitem[{Zheng et~al.(2018)Zheng, Stapleton, Fox, and Dunne}]{ZHENG2018}
\bibinfo{author}{Z.~Zheng}, \bibinfo{author}{A.~Stapleton},
  \bibinfo{author}{K.~Fox}, \bibinfo{author}{F.~P. Dunne},
\newblock \bibinfo{title}{Understanding thermal alleviation in cold dwell
  fatigue in titanium alloys},
\newblock \bibinfo{journal}{International Journal of Plasticity}
  \bibinfo{volume}{111} (\bibinfo{year}{2018}) \bibinfo{pages}{234 -- 252}.
  \URLprefix
  \url{http://www.sciencedirect.com/science/article/pii/S0749641918302195}.
  \DOIprefix\doi{https://doi.org/10.1016/j.ijplas.2018.07.018}.
\bibitem[{Guery et~al.(2016)Guery, Hild, Latourte, and Roux}]{GUERY2016}
\bibinfo{author}{A.~Guery}, \bibinfo{author}{F.~Hild},
  \bibinfo{author}{F.~Latourte}, \bibinfo{author}{S.~Roux},
\newblock \bibinfo{title}{Slip activities in polycrystals determined by
  coupling dic measurements with crystal plasticity calculations},
\newblock \bibinfo{journal}{International Journal of Plasticity}
  \bibinfo{volume}{81} (\bibinfo{year}{2016}) \bibinfo{pages}{249 -- 266}.
  \URLprefix
  \url{http://www.sciencedirect.com/science/article/pii/S0749641916000188}.
  \DOIprefix\doi{https://doi.org/10.1016/j.ijplas.2016.01.008}.
\bibitem[{Stroh and Mott(1954)}]{Stroh1954}
\bibinfo{author}{A.~N. Stroh}, \bibinfo{author}{N.~F. Mott},
\newblock \bibinfo{title}{The formation of cracks as a result of plastic flow},
\newblock \bibinfo{journal}{Proceedings of the Royal Society of London. Series
  A. Mathematical and Physical Sciences} \bibinfo{volume}{223}
  (\bibinfo{year}{1954}) \bibinfo{pages}{404--414}. \URLprefix
  \url{https://royalsocietypublishing.org/doi/abs/10.1098/rspa.1954.0124}.
  \DOIprefix\doi{10.1098/rspa.1954.0124}.
\bibitem[{Britton et~al.(2015)Britton, Dunne, and Wilkinson}]{BRITTON2015}
\bibinfo{author}{T.~B. Britton}, \bibinfo{author}{F.~P.~E. Dunne},
  \bibinfo{author}{A.~J. Wilkinson},
\newblock \bibinfo{title}{On the mechanistic basis of deformation at the
  microscale in hexagonal close-packed metals},
\newblock \bibinfo{journal}{Proceedings of the Royal Society A: Mathematical,
  Physical and Engineering Sciences} \bibinfo{volume}{471}
  (\bibinfo{year}{2015}) \bibinfo{pages}{20140881}. \URLprefix
  \url{https://royalsocietypublishing.org/doi/abs/10.1098/rspa.2014.0881}.
  \DOIprefix\doi{10.1098/rspa.2014.0881}.
  \href{http://arxiv.org/abs/https://royalsocietypublishing.org/doi/pdf/10.1098/rspa.2014.0881}{{\tt
  arXiv:https://royalsocietypublishing.org/doi/pdf/10.1098/rspa.2014.0881}}.
\bibitem[{Sakai and Fine(1974{\natexlab{a}})}]{SAKAI19741359}
\bibinfo{author}{T.~Sakai}, \bibinfo{author}{M.~Fine},
\newblock \bibinfo{title}{Plastic deformation of ti-al single crystals in
  prismatic slip},
\newblock \bibinfo{journal}{Acta Metallurgica} \bibinfo{volume}{22}
  (\bibinfo{year}{1974}{\natexlab{a}}) \bibinfo{pages}{1359 -- 1372}.
  \URLprefix
  \url{http://www.sciencedirect.com/science/article/pii/0001616074900364}.
  \DOIprefix\doi{https://doi.org/10.1016/0001-6160(74)90036-4}.
\bibitem[{Sakai and Fine(1974{\natexlab{b}})}]{SAKAI1974545}
\bibinfo{author}{T.~Sakai}, \bibinfo{author}{M.~Fine},
\newblock \bibinfo{title}{Basal slip of ti-al single crystals},
\newblock \bibinfo{journal}{Scripta Metallurgica} \bibinfo{volume}{8}
  (\bibinfo{year}{1974}{\natexlab{b}}) \bibinfo{pages}{545 -- 547}. \URLprefix
  \url{http://www.sciencedirect.com/science/article/pii/0036974874900660}.
  \DOIprefix\doi{https://doi.org/10.1016/0036-9748(74)90066-0}.
\bibitem[{Sakai and Fine(1974{\natexlab{c}})}]{Sakai1974FailureOS}
\bibinfo{author}{T.~Sakai}, \bibinfo{author}{M.~Fine},
\newblock \bibinfo{title}{Failure of schmid's law in tial alloys for prismatic
  slip},
\newblock \bibinfo{journal}{Scripta Metallurgica} \bibinfo{volume}{8}
  (\bibinfo{year}{1974}{\natexlab{c}}) \bibinfo{pages}{541--544}.
\bibitem[{Bache et~al.(2010)Bache, Dunne, and Madrigal}]{BACHE2010}
\bibinfo{author}{M.~R. Bache}, \bibinfo{author}{F.~P.~E. Dunne},
  \bibinfo{author}{C.~Madrigal},
\newblock \bibinfo{title}{Experimental and crystal plasticity studies of
  deformation and crack nucleation in a titanium alloy},
\newblock \bibinfo{journal}{The Journal of Strain Analysis for Engineering
  Design} \bibinfo{volume}{45} (\bibinfo{year}{2010})
  \bibinfo{pages}{391--399}. \URLprefix
  \url{https://doi.org/10.1243/03093247JSA594}.
  \DOIprefix\doi{10.1243/03093247JSA594}.
\bibitem[{Germain et~al.(2008)Germain, Gey, Humbert, Vo, Jahazi, and
  Bocher}]{Germain2008}
\bibinfo{author}{L.~Germain}, \bibinfo{author}{N.~Gey},
  \bibinfo{author}{M.~Humbert}, \bibinfo{author}{P.~Vo},
  \bibinfo{author}{M.~Jahazi}, \bibinfo{author}{P.~Bocher},
\newblock \bibinfo{title}{Texture heterogeneities induced by subtransus
  processing of near $\alpha$ titanium alloys},
\newblock \bibinfo{journal}{Acta Materialia} \bibinfo{volume}{56}
  (\bibinfo{year}{2008}) \bibinfo{pages}{4298--4308}.
  \DOIprefix\doi{10.1016/j.actamat.2008.04.065}.
\bibitem[{Le~Biavant et~al.(2002)Le~Biavant, Pommier, and Prioul}]{BIAVANT2002}
\bibinfo{author}{K.~Le~Biavant}, \bibinfo{author}{S.~Pommier},
  \bibinfo{author}{C.~Prioul},
\newblock \bibinfo{title}{Local texture and fatigue crack initiation in a
  ti-6al-4v titanium alloy},
\newblock \bibinfo{journal}{Fatigue \& Fracture of Engineering Materials \&
  Structures} \bibinfo{volume}{25} (\bibinfo{year}{2002})
  \bibinfo{pages}{527--545}. \URLprefix
  \url{https://onlinelibrary.wiley.com/doi/abs/10.1046/j.1460-2695.2002.00480.x}.
  \DOIprefix\doi{https://doi.org/10.1046/j.1460-2695.2002.00480.x}.
\bibitem[{Bantounas et~al.(2010)Bantounas, Dye, and Lindley}]{BANTOUNAS2010}
\bibinfo{author}{I.~Bantounas}, \bibinfo{author}{D.~Dye},
  \bibinfo{author}{T.~C. Lindley},
\newblock \bibinfo{title}{The role of microtexture on the faceted fracture
  morphology in ti–6al–4v subjected to high-cycle fatigue},
\newblock \bibinfo{journal}{Acta Materialia} \bibinfo{volume}{58}
  (\bibinfo{year}{2010}) \bibinfo{pages}{3908 -- 3918}. \URLprefix
  \url{http://www.sciencedirect.com/science/article/pii/S1359645410001916}.
  \DOIprefix\doi{https://doi.org/10.1016/j.actamat.2010.03.036}.
\bibitem[{Zhang et~al.(2015)Zhang, Yang, Huang, Wu, and Davies}]{ZHANGK2015}
\bibinfo{author}{K.~Zhang}, \bibinfo{author}{K.~Yang},
  \bibinfo{author}{A.~Huang}, \bibinfo{author}{X.~Wu},
  \bibinfo{author}{C.~Davies},
\newblock \bibinfo{title}{Fatigue crack initiation in as forged ti–6al–4v
  bars with macrozones present},
\newblock \bibinfo{journal}{International Journal of Fatigue}
  \bibinfo{volume}{80} (\bibinfo{year}{2015}) \bibinfo{pages}{288 -- 297}.
  \URLprefix
  \url{http://www.sciencedirect.com/science/article/pii/S0142112315001772}.
  \DOIprefix\doi{https://doi.org/10.1016/j.ijfatigue.2015.05.020}.
\bibitem[{Zhang et~al.(2017)Zhang, Yang, Lim, Wu, and Davies}]{ZHANGK2017}
\bibinfo{author}{K.~Zhang}, \bibinfo{author}{K.~Yang},
  \bibinfo{author}{S.~Lim}, \bibinfo{author}{X.~Wu},
  \bibinfo{author}{C.~Davies},
\newblock \bibinfo{title}{Effect of the presence of macrozones on short crack
  propagation in forged two-phase titanium alloys},
\newblock \bibinfo{journal}{International Journal of Fatigue}
  \bibinfo{volume}{104} (\bibinfo{year}{2017}) \bibinfo{pages}{1 -- 11}.
  \URLprefix
  \url{http://www.sciencedirect.com/science/article/pii/S014211231730292X}.
  \DOIprefix\doi{https://doi.org/10.1016/j.ijfatigue.2017.07.002}.
\bibitem[{Jun et~al.(2016)Jun, Armstrong, and Britton}]{JUN2016NANO}
\bibinfo{author}{T.-S. Jun}, \bibinfo{author}{D.~E. Armstrong},
  \bibinfo{author}{T.~B. Britton},
\newblock \bibinfo{title}{A nanoindentation investigation of local strain rate
  sensitivity in dual-phase ti alloys},
\newblock \bibinfo{journal}{Journal of Alloys and Compounds}
  \bibinfo{volume}{672} (\bibinfo{year}{2016}) \bibinfo{pages}{282 -- 291}.
  \URLprefix
  \url{http://www.sciencedirect.com/science/article/pii/S0925838816304133}.
  \DOIprefix\doi{https://doi.org/10.1016/j.jallcom.2016.02.146}.
\bibitem[{Xiong et~al.(2020)Xiong, Karamched, Nguyen, Collins, Magazzeni,
  Tarleton, and Wilkinson}]{XIONG2020}
\bibinfo{author}{Y.~Xiong}, \bibinfo{author}{P.~S. Karamched},
  \bibinfo{author}{C.-T. Nguyen}, \bibinfo{author}{D.~M. Collins},
  \bibinfo{author}{C.~M. Magazzeni}, \bibinfo{author}{E.~Tarleton},
  \bibinfo{author}{A.~J. Wilkinson},
\newblock \bibinfo{title}{Cold creep of titanium: Analysis of stress relaxation
  using synchrotron diffraction and crystal plasticity simulations},
\newblock \bibinfo{journal}{Acta Materialia} \bibinfo{volume}{199}
  (\bibinfo{year}{2020}) \bibinfo{pages}{561 -- 577}. \URLprefix
  \url{http://www.sciencedirect.com/science/article/pii/S1359645420306078}.
  \DOIprefix\doi{https://doi.org/10.1016/j.actamat.2020.08.010}.
\bibitem[{Jun et~al.(2016)Jun, Zhang, Sernicola, Dunne, and Britton}]{JUN2016}
\bibinfo{author}{T.-S. Jun}, \bibinfo{author}{Z.~Zhang},
  \bibinfo{author}{G.~Sernicola}, \bibinfo{author}{F.~P. Dunne},
  \bibinfo{author}{T.~B. Britton},
\newblock \bibinfo{title}{Local strain rate sensitivity of single $\alpha$
  phase within a dual-phase ti alloy},
\newblock \bibinfo{journal}{Acta Materialia} \bibinfo{volume}{107}
  (\bibinfo{year}{2016}) \bibinfo{pages}{298 -- 309}. \URLprefix
  \url{http://www.sciencedirect.com/science/article/pii/S1359645416300544}.
  \DOIprefix\doi{https://doi.org/10.1016/j.actamat.2016.01.057}.
\bibitem[{Tanaka and Higashida(2016)}]{TANAKA2016}
\bibinfo{author}{M.~Tanaka}, \bibinfo{author}{K.~Higashida},
\newblock \bibinfo{title}{Temperature dependence of effective stress in
  severely deformed ultralow-carbon steel},
\newblock \bibinfo{journal}{Philosophical Magazine} \bibinfo{volume}{96}
  (\bibinfo{year}{2016}) \bibinfo{pages}{1979--1992}. \URLprefix
  \url{https://doi.org/10.1080/14786435.2016.1183828}.
  \DOIprefix\doi{10.1080/14786435.2016.1183828}.
  \href{http://arxiv.org/abs/https://doi.org/10.1080/14786435.2016.1183828}{{\tt
  arXiv:https://doi.org/10.1080/14786435.2016.1183828}}.
\bibitem[{Zheng et~al.(2017)Zheng, Balint, and Dunne}]{ZHENG2017Mechanistic}
\bibinfo{author}{Z.~Zheng}, \bibinfo{author}{D.~S. Balint},
  \bibinfo{author}{F.~P. Dunne},
\newblock \bibinfo{title}{Mechanistic basis of temperature-dependent dwell
  fatigue in titanium alloys},
\newblock \bibinfo{journal}{Journal of the Mechanics and Physics of Solids}
  \bibinfo{volume}{107} (\bibinfo{year}{2017}) \bibinfo{pages}{185 -- 203}.
  \URLprefix
  \url{http://www.sciencedirect.com/science/article/pii/S0022509616307918}.
  \DOIprefix\doi{https://doi.org/10.1016/j.jmps.2017.07.010}.
\bibitem[{Ozaki et~al.(2017)Ozaki, Yano, and Onishi}]{Ozaki2017}
\bibinfo{author}{Y.~Ozaki}, \bibinfo{author}{M.~Yano},
  \bibinfo{author}{M.~Onishi},
\newblock \bibinfo{title}{Effective hyperparameter optimization using
  nelder-mead method in deep learning},
\newblock \bibinfo{journal}{IPSJ Transactions on Computer Vision and
  Applications} \bibinfo{volume}{9} (\bibinfo{year}{2017}) \bibinfo{pages}{20}.
  \URLprefix \url{https://doi.org/10.1186/s41074-017-0030-7}.
  \DOIprefix\doi{10.1186/s41074-017-0030-7}.

\end{thebibliography}

\end{document}